\documentclass[aps,a4paper,pre,eqsecnum,showpacs,floatfix,twocolumn,merge,longbibliography,elide]{revtex4-1}

\usepackage{amsmath}
\usepackage[usenames,dvipsnames,svgnames]{pstricks}		% more colors

\usepackage{graphicx}		% including|inputing graphics 
\usepackage{tikz}			% drawning graphics
\usepackage[charter,expert]{mathdesign}				% new mathmode

\usepackage[ulem=normalem]{changes}
\setaddedmarkup{\blue{#1}}
\setdeletedmarkup{\red\sout{#1}}

\newcommand{\REM}[1]{}

\usepackage[%
	  unicode=true,%
	  colorlinks=true,%
	  linkcolor=blue,anchorcolor=blue,%
	  breaklinks=true,%
	  citecolor=blue,filecolor=cyan,%
	  menucolor=blue,%
	  breaklinks=true,%
	  urlcolor=blue]{hyperref}
	
\begin{document}
	
\title{ A systematic approach to determine the spectral
characteristics of molecular magnets
	}
\author{M.~Georgiev}
\email{mgeorgiev@issp.bas.bg}
\affiliation{Institute of Solid State Physics, Bulgarian Academy of Sciences,
Tsarigradsko Chauss\'ee 72, 1784 Sofia, Bulgaria}
\author{H.~Chamati}
\affiliation{Institute of Solid State Physics, Bulgarian Academy of Sciences,
Tsarigradsko Chauss\'ee 72, 1784 Sofia, Bulgaria}

\date{\today}
\begin{abstract} 
We devise a formalism to investigate in a systematic
way the spectroscopic magnetic
excitations in molecular magnets.
This consists in introducing a bilinear spin Hamiltonian that allows
for discrete
coupling parameters accounting for distinct spin coupling mechanisms 
among the constituent magnetic ions,
as well as the influence of the nonmagnetic ions in the system.
The model is applied to explore the magnetic excitations of the
trimeric magnetic compounds
$\mathrm{A_3Cu_3(PO_4)_4}$ \
$\mathrm{(A\ =\ Ca,\ Sr,\ Pb)}$ and the tetrameric molecular magnet $\mathrm{Ni_4Mo_{12}}$. 
Our results are in a very good
agreement with the available experimental data:
For all trimers $\mathrm{A_3Cu_3(PO_4)_4}$, calculations reveal the existence of one
thin energy band referring to the flatness of observed excitation peaks. 
Moreover for the tetramer $\mathrm{Ni_4Mo_{12}}$, we concluded that the magnetic
excitations may be traced back to the
specific geometry and complex chemical structure of the exchange bridges
leading to the splitting and broadness of the peaks centered about 0.5
meV and 1.7 meV.
\end{abstract}
\maketitle

\section{Introduction}
Molecular nanomagnets have seen a resurgence of interest in recent
years (for an extensive review see e.g. Ref.
\cite{sieklucka_molecular_2017} and references therein). Their
small size allows precise characterizability
both theoretically and experimentally. They possess unique properties
and are ideal candidates for exploring the interplay of the quantum and
the classical worlds.
Their magnetic properties are determined from the
collective behavior of weakly interacting fundamental structural
units forming isolated dimers, trimers and tetramers
\cite{furrer_magnetic_2013}.
They have great potential for technological applications:
%in information storage and quantum computation.
The effect of quantum tunnelling in single-molecule magnets
\cite{wernsdorfer_exchange-biased_2002,schenker_phonon_2005}, the
response of spin-switching in the frustrated antiferromagnetic
chromium trimmer \cite{jamneala_kondo_2001} and even-odd effects in spin chain magnets
\cite{machens_even-odd_2013} are some prominent examples.
Furthermore, the molecular magnet $\mathrm{Ni_4Mo_{12}}$ provides a
unique opportunity for exploring unusual magnetic behavior
\cite{schnack_observation_2006,kostyuchenko_non-heisenberg_2007},
while the difference in the magnetic properties \cite{ghosh_magnetic_2010}
among the compounds $\mathrm{Ca_3Cu_2Ni(PO_4)_4}$ and
$\mathrm{Ca_3Cu_2Mg(PO_4)_4}$ shows the richness of the physical features of linear
spin trimers (see e.g. \cite{ghosh_nmr_2010,ghosh_spin_2012}).
It is worth mentioning that even the structure of the nucleon and the
distribution of its spin degrees of freedom are
not yet fully understood \cite{aidala-spin-2013} signalling the
continuous scientific interest in exploring the features of the ``smallest'' quantum spin systems.
Whether
in nuclear physics or in solids these systems play an important role
for testing theoretical formalisms. The nature of the underlying quantum
collective processes such as higher order spin exchange interactions
can be analysed in terms of different spin Hamiltonians
\cite{coldea_spin_2001,zaharko_tetrahedra_2008,islam_first-principles_2010,klemm_single-ion_2008,gatteschi_epr_2006}.
Within the nature of spin exchange processes nanomagnets can be
studied also in the context of quantum estimation theory
\cite{troiani_probing_2016}. Theoretical analysis of molecular magnets
$\mathrm{Fe_8}$ and $\mathrm{Mn_{12}}$ based on the Grover algorithm
\cite{leuenberger_quantum_2001}, makes them promising candidates
for building memory devices. Moreover the heterometallic linkers $\mathrm{Cr_7Ni}$
molecular rings were used to investigate the propagation of spin
information at the supramolecular scale \cite{bellini_propagation_2011}.

Magnetic molecules possess intrinsic properties and are ideal systems
to gain useful insights into the underlying coupling mechanisms. On the experimental
side, Inelastic Neutron Scattering (INS) 
\cite{lovesey_1986,malcolm_neutron_1989,furrer_neutron_2009,toperverg_neutron_2015} 
plays a central role in
determining the exchange effects and relevant magnetic spectra. In
complement to different magnetic measurment methods, INS techniques appear
to be of high value, and in the past decades it has been widely
applied to explore the properties of spin clusters. INS experiments on
the spin dimer
$[\text{Ni}_2(\text{ND}_2\text{C}_2\text{H}_4\text{ND}_2)_4\text{Br}_2]\text{Br}_2$
has demonstrated the important
contribution of neutron spectroscopy \cite{stebler_intra_1982}. INS measurments were obtained
for different magnetic clusters, such as: The trimer
$\mathrm{La_4Cu_3MoO_{12}}$, with strong intratrimer antiferromagnetic
interactions, where the copper ions form an isolated triangle
\cite{azuma_antiferromagnetic_2000}, the dimer
$\text{SrCu}_2(\text{BO}_3)_2$ with observed multiplet excitations
\cite{kageyama_direct_2000,gaulin_high-resolution_2004} the
polyoxomolybdate $\text{Mn}_{72}\text{Fe}_{32}$
\cite{garlea_probing_2006}, and the magnetic molecule Fe$_9$ in presence
of an external magnetic field \cite{vaknin_magnetic_2014}.

The physical properties, such as energy spectra, susceptibility,
etc., of magnetic clusters at the nanoscale depend
on their size, shape (for more details see
\cite{chamati_theory_2013,sellmyer_novel_2015,sieklucka_molecular_2017}
and references
therein) and the presence of different bondings among the
constituent chemical elements.
Thus the distribution of ligands with different strength in
conjunction with finite-size, 
as well as surface effects 
have huge impact on their
characteristics.

When studying the spectral properties of single magnetic
molecules, usually
(see e.g. Ref. \cite{furrer_magnetic_2013} and references therein)
one relies on the structural symmetry of the cluster to solve
the ensuing quantum mechanical problem.
Thus grouping symmetrically
equivalent spins into a resulting single one employing the sum rules of angular
momenta. Further one adds more spins according to well defined
selection rules till fully characterizing the specific cluster under
consideration.

The main aim of the present paper is to
propose an alternative approach that leads naturally to computing
the relevant physical quantities of any magnetic cluster.
Furthermore, it is able to reproduce reasonably well the experimental 
results. The present approach is based on the assumption that 
in a molecular magnet with nontrivial geometry and complex chemical environment
neither the exchange path between two magnetic ions nor the corresponding coupling  
are unique.
This causes the transition energy to vary, leading to a broadened excitation 
width in the energy spectrum and even splitting.
Accordingly the number of all energy values form a set
that can uniquely 
identify the most relevant bonds, despite being identical to each
other.
The associated effect could be 
studied by accounting for appropriate spin coupling parameters.
To this end, we introduce a bilinear microscopic spin Hamiltonian with discrete couplings 
that allow for distinct spin coupling mechanisms among equivalent spins 
allowing one to identify the different exchange paths. 
Hence, one can precisely determine the energy levels and
the relevant magnetic characteristics of magnetic clusters and the underlying
physical processes of experimentally observed spectra.

The present method is powerful and quite
general. It can be applied to a variety of
physical problems, such as unveiling the structure of the nucleus
(see e.g. Ref. \cite{thomas_interplay_2008}). Here, it
will be tested on two classes of molecular magnets that have generated a great
deal of interest by many researchers both on the theoretical as well as
the experimental sides.

\begin{figure}[h!]
\vspace*{0.2cm}
\includegraphics[scale=0.8]{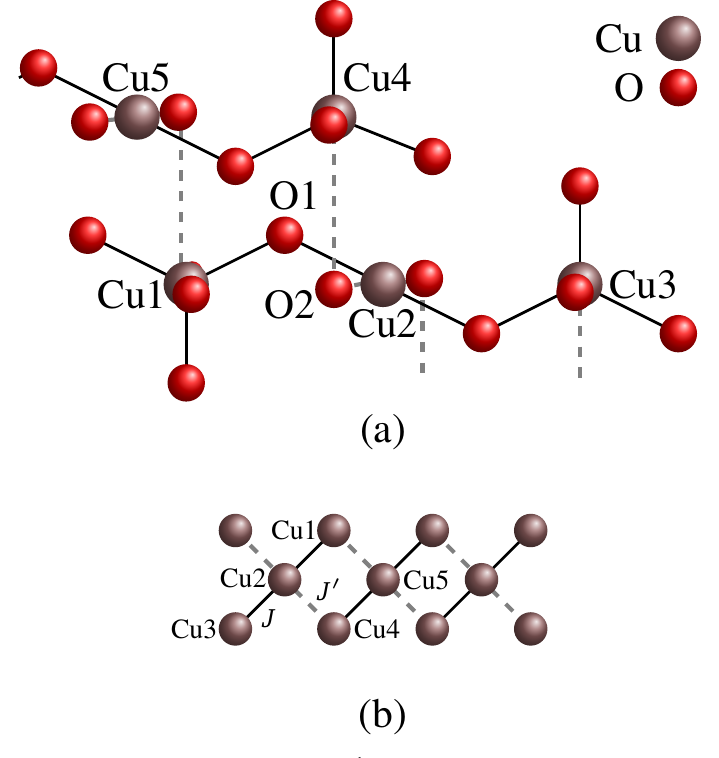}
\caption{(a)
Exchange pathways in $\mathrm{A_3Cu_3(PO_4)_4}$ $(\mathrm{A = Ca, Sr, Pb})$.
Copper colored circles represent copper ions,
the red ones stand for oxygen atoms. The solid (black) and dashed
(gray) lines represent the intratrimer and intertrimer exchange pathways,
respectively. (b) Schematic representation of the intratrimer $J$ and
intertrimer $J'$ magnetic interactions in the array of isolated trimers.
\label{fig:Copper}}
\end{figure}

The first class of materials that are the focus of our attention belongs to the
family of compounds $\mathrm{A_3Cu_3(PO_4)_4}$ with $\mathrm{(A = Ca,
Sr, Pb)}$, where the three spin-half $\mathrm{Cu^{2+}}$ ions form a linear
trimer (see FIG. \ref{fig:Copper}). 
Magnetic measurements on trimer copper chains with
$\mathrm{(A = Ca, Sr)}$ are reported in Ref. \cite{drillon_1d_1993}
and analysed in the framework of Heisenberg and Ising models.
It was shown that the intertrimer interactions are negligible and thus the
trimers might be considered as separate clusters.
These results were confirmed
via INS experiments \cite{matsuda_magnetic_2005,podlesnyak_magnetic_2007}
that shed light on the magnetic
spectra with the aid of the antiferromagnetic Heisenberg model
involving nearest and next-nearest intratrimer interactions, and
later they were extended to the
compound $\mathrm{Ca_3Cu_3(PO_4)_4}$
\cite{furrer_magnetic_2010}. Moreover, it turns
out that the interaction between edge spins in the isolated trimer is
also negligible.

\begin{figure}[!ht]
\vspace*{0.2cm}
\includegraphics[scale=0.8]{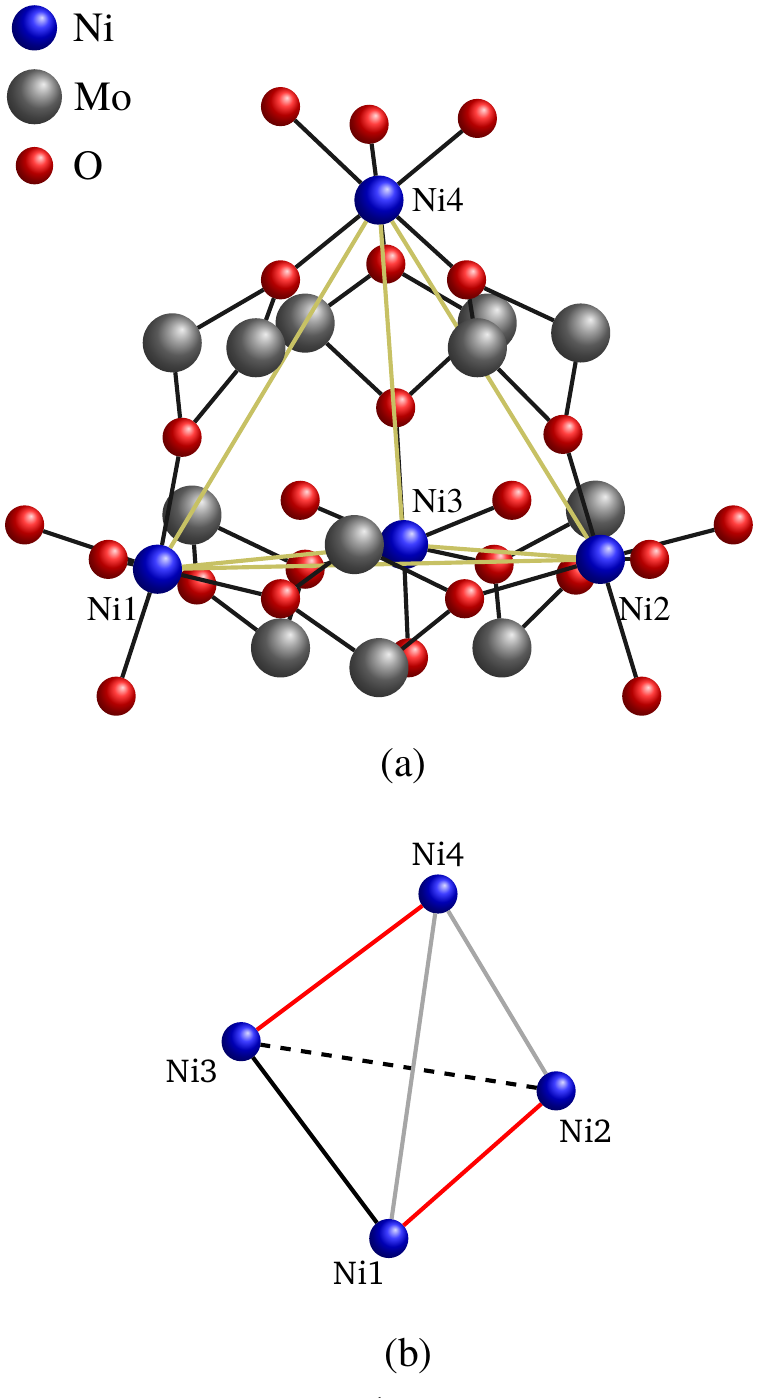}
\caption{
(a) Sketch of the structure of the molecular
nanomagnet Ni$_4$Mo$_{12}$. (b) Schematic view of the arrangment of Ni
ions (blue balls). The grey
lines represent the two shorter distances, while the red lines show the
spin-1 dimers. \label{fig:Nickelconf}}
\end{figure}

The second material of interest is the magnetic molecule
$[\mathrm{Mo_{12}O_{30}(\mu_2-OH)_{10}H_2(Ni(H_2O)_3)_4}]$, denoted by
$\mathrm{Ni_4Mo_{12}}$, where four spin-1 $\mathrm{Ni^{2+}}$ ions are
sitting on the vertices of a distorted tetrahedron (see FIG.
\ref {fig:Nickelconf}). This molecule shows an unusual magnetic behavior
\cite{schnack_observation_2006}.
It was suggested \cite{kostyuchenko_non-heisenberg_2007} that the
experimental data could be explained by accounting for a three
interaction term in addition to the Heisenberg nearest-neighbor
exchange and a biquadratic term.
%This spin Hamiltonian turns out to be
%consistent with experimentally observed magnetization steps.
The theoretical description of INS data, especially the
intensity and the width of the peak at about $1.7 \
\textrm{meV}$ has attracted lot of interest (for more details see Ref.
\cite{nehrkorn_inelastic_2010} and references therein).
%Employing
%a preselected spin coupling scheme Ref.
In Ref. \cite{furrer_magnetic_PRB_2010} it was pointed out
that the Heisenberg model with single-ion anisotropy is a Hamiltonian adequate to
reproduce the main features of experimentally obtained INS data.
However, even by including
higher order terms and/or perturbations, such as
single-ion anisotropy an accurate reproduction of the
experimental INS spectrum
is not yet reported.
	
The rest of this paper is structured as follows: In Section
\ref{sec:fundamprinc} we
present the details of our approach and its advantages when applied to spin systems. We formulate
explicitly the
Hamiltonian and the key constraints that allow the derivation of
the main results throughout the rest of paper.
In Sections \ref{sec:copper} and \ref{sec:nickel}
we explore the low-lying magnetic excitations of the
compounds $\mathrm{A_3Cu_3(PO_4)_4}$ (where A stands for Ca, Sr, Pb) and
$\mathrm{Ni_4Mo_{12}}$.
A summary of the results obtained throughout this paper are
presented in Section \ref{sec:conclusion}.

\section{The model and the method}\label{sec:fundamprinc}
\subsection{INS and the Heisenberg model}
The study of magnetic excitations determined by
INS techniques on one hand requires a specific microscopic model, and on
the other an analysis of the neutron scattering probabilities
\cite{lovesey_1986,malcolm_neutron_1989,furrer_neutron_2009,toperverg_neutron_2015}.
To determine
the energy level structure and the transitions corresponding to
the experimentally observed magnetic spectra one needs a minimal
number of parameters to account for all couplings in the system. 
It is cumbersome
to apply a general approach with a unique set of parameters
that can describe all possible magnetic effects and in addition to
distinguish between inter-molecular and intra-molecular features. 
The principal assumption of our method is that
the magnetic excitations of spin clusters obtained by INS are mainly governed by the
exchange of electrons between
the constituent ions. Then, the experimental data are interpreted in terms
of a well defined microscopic model.
In the absence of anisotropy, i.e. negligible spin-orbit coupling, 
the exchange interaction in molecular magnets can be described by 
the Heisenberg model
\begin{equation}\label{eq:MainHamiltonian}
\hat{H} = \sum\limits_{i \ne j}^{} J_{ij}
\hat{\mathbf{s}}_i \cdot \hat{\mathbf{s}}_j,
\end{equation}	
where $J_{ij} = J_{ji}$ is the exchange coupling that effectively accounts
for the electrostatic interaction between the $i$th and $j$th ions and
represents the amount of transition energy arising due to the electron's spins.
Hamiltonian \eqref{eq:MainHamiltonian} commutes with the square and each component of the total spin 
operator $\hat{\mathbf{s}}=\sum_i \hat{\mathbf{s}}_i$.
Therefore the eigenvalues of Hamiltonian \eqref{eq:MainHamiltonian} can be computed 
within the total spin operator eigenstates $\rvert s,m \rangle$, where $s$ 
and $m$ stand for the total spin and magnetic quantum numbers, respectively.

Depending on the geometry of the specific cluster under consideration, other magnetic and
non-magnetic properties may be taken into account by generalizing the
Hamiltonian \eqref{eq:MainHamiltonian}. The general practice is
to include different interaction terms 
referring to the type of exchange under consideration. Such terms are
biquadratic
\cite{blume_biquadratic_1969,lacroix_spin_2011,smerald_theory_2013},
four-spin
\cite{ivanov_diamond_2009,muller_perturbation_2002,lauchli_two-step_2005},
three-body \cite{ivanov_phase_2014,michaud_phase_2013} or high order
multipolar interaction terms \cite{santini_multipolar_2009}. Even with
some of the aforementioned interactions the eigenvalues of the ensuing
Hamiltonian may remain degenerates with respect to the total magnetic
quantum number, leaving the experimentally observed splitting effects of the magnetic
spectrum unexplained. Furthermore, the interplay between the different terms may break the
rotational symmetry and the total spin $s$ may no longer be a good quantum number.
Moreover one may include perturbation terms like the single-ion anisotropy which
arises due to the one site spin-orbit coupling \cite{rudowicz_2015}.

The identification of the experimentally observed magnetic peaks in the
obtained energy level structure, estimated by the considered
microscopic model, is not unique. To obtain meaningful results one has
to calculate the scattering intensities $I_{n'n}(\mathbf{q})$,
integrated over the angles of the scattering vector $\mathbf{q}$,
of the existing transitions and analyse their
dependence on the temperature and the magnitude of the neutron scattering vector.
For identical magnetic ions, we have
\cite{lovesey_1986,malcolm_neutron_1989,furrer_neutron_2009,toperverg_neutron_2015}
\begin{equation}\label{eq:ScateringIntensities}
I_{n'n}(\mathbf{q}) \propto F^2(\mathrm{q}) \sum_{\alpha, \beta} \varTheta^{\alpha \beta}
S^{\alpha \beta} (\mathbf{q},\omega_{n'n}).
\end{equation}
Here $\mathbf{q} = \mathbf{k}_0 - \mathbf{k}$,
with $\mathbf{k}_0$ and $\mathbf{k}$ -- the incoming
and the scattered neutron wave vectors, respectively. The magnitudes of
these vectors are denoted by $\mathrm{q}$, $k_0$ and $k$. The
transition's frequency, with neutron's mass $\mu$, is given by $\omega =
(2\mu)^{-1}(\mathbf{k}^2-\mathbf{k}^2_0)$, $F(\mathrm{q})$ -- the
spin magnetic form factor,
$\varTheta^{\alpha \beta}$ is the polarization factor, 
and $\alpha, \beta, \gamma \in \{x,y,z\}$. In \eqref{eq:ScateringIntensities} the magnetic
scattering functions are explicitly written as
\begin{align}\label{eq:ScatterFunctions}
S^{\alpha \beta} (\mathbf{q},\omega) = & \ \sum_{n,n',i,j}^{} e^{\mathrm{i} \mathbf{q} \cdot \mathbf{r}_{ij}} p_n \langle n \lvert \hat{s}^\alpha_i \rvert n' \rangle \langle n'\rvert \hat{s}^\beta_j \rvert n \rangle \, \delta(\hbar \omega - E_{n'n}),
 \\
& \quad p_n = Z^{-1} e^{-\frac{E_n}{k_B T}}, \nonumber
\end{align}
where $\rvert n \rangle$, $\rvert n' \rangle$ are the initial and
final states with the corresponding energy $E_n$ and $E_{n'}$,
respectively, $E_{n'n} (= E_{n'}-E_{n})$ -- the transition energy
and $Z$ is the partition function.
The term $e^{\mathrm{i} \mathbf{q} \cdot
\mathbf{r}_{ij}}$ is the structure factor associated with the cluster
geometry. 
When $s$ is a good quantum number  the eigenstates $\rvert n \rangle \equiv\
\rvert s,m \rangle$ and $\rvert n' \rangle \equiv\ \rvert s',m'
\rangle$, where $s$ and $m$ stand for the
total spin and magnetic quantum numbers, respectively.
Therefore, a magnetic transition sets in when $\omega \equiv
\omega_{n'n}$.

The spin magnetic form factor \cite{jensen_rare_1991}
is given by
\begin{equation}\label{eq:MagFormFactor}
F(\mathrm{q}) = \int\limits_{0}^{\infty} \mathrm{r}^2 R^2_{i0}(\mathrm{r}) J_{\nu}(\mathrm{q},\mathrm{r}) \mathrm{d} \mathrm{r}, 
\end{equation}
where $R_{i0}(\mathrm{r})$ are the radial wave functions and
$J_{\nu}(\mathrm{q},\mathrm{r})$ are the spherical Bessel functions of
the first kind. The advantage of INS is that one can clearly
distinguish the magnetic transitions from phonon excitations, as
the former obey different statistics and decrease by increasing the magnitude
of scattering vector. Furthermore, this method does not require an
external magnetic field, since the neutron spin interacts with the
intrinsic magnetic field of the cluster.

\subsection{Phenomenological spin model} \label{sec:spinmodel}
In molecular magnets, the distribution 
of coupled spins (dimers) plays a crucial
role in uniquely determining the scattering intensities.
Even when the bonds are indistinguishable with respect to
their lengths and the total spin of the coupled spins, according to
\eqref{eq:ScatterFunctions}, one can clearly obtain different in 
magnitude neutron scattering intensities. However, to distinguish
the intensities one has to use an appropriate
spin model leading to an energy sequence such that the 
$\delta$ function in the r.h.s of \eqref{eq:ScatterFunctions} 
identifies the spin bonds 
with respect to the structure factors. 
Notice that, even with a selected \textit{a priori} spin coupling scheme,
the Hamiltonian \eqref{eq:MainHamiltonian} may not be adequate to 
obtain the correct energy structure.

In the quest of a procedure that allows to characterize uniquely
each bond in a magnetic cluster assuming
nonuniqueness of exchange pathways
we propose the following Hamiltonian
\begin{equation}\label{eq:AddHamiltonian}
\hat{\mathcal{H}} = 
\sum\limits_{i \ne j}^{} J_{ij} 
\hat{\boldsymbol{\sigma}}_i \cdot \hat{\mathbf{s}}_j,
\end{equation}
where the couplings $J_{ij} = J_{ji}$
are effective exchange constants and the operator
$\hat{\boldsymbol{\sigma}}_i \equiv
(\hat{\sigma}^x_i, \hat{\sigma}^y_i, \hat{\sigma}^z_i)$ 
accounts for the differences in local coupling processes of the $i$-th ion.
If $J_{ij}$ is not indentical for all pairs $i$ and $j$, then the sigma operators will  
differ from their associated spin operators. This will allow one to obtain the whole set of 
transition energies corresponding to the exchange between $i$th and $j$th ions.

For a single spin the square and $z$ component of each operator
$\sigma$ are completely determined in the
basis of the total spin component $\hat{s}^z$, such that for all $i$ and 
$\alpha \in \{x,y,z\}$
\begin{equation}\label{eq:Sigma_i}
\hat{\sigma}^\alpha_i \lvert \ldots ,s_i,m_i, \ldots \rangle
	=
a^{s_i,m_i}_i \hat{s}^{\alpha_{\phantom{i}}}_i 
\lvert \ldots ,s_i,m_i, \ldots \rangle,
\end{equation}
where $a^{s_i,m_i}_i \in \mathbb{R}$. %are non zero scalars.
Furthermore, the $\sigma$ rising and lowering operators
obey the equations
\begin{equation}\label{eq:Sigma_LR_i}
\hat{\sigma}^{\pm}_i \lvert \ldots ,s_i,m_i, \ldots \rangle
=
a^{s_i,m_i}_i \hat{s}^{{\pm}_{\phantom{i}}}_i 
\lvert \ldots ,s_i,m_i, \ldots \rangle.
\end{equation}
For all $i$, the square of $\sigma_i$ commutes
only with its $z$ component. 
Its eigenvalues depend on $m_i$ and according to
\eqref{eq:Sigma_i} and \eqref{eq:Sigma_LR_i} 
one can distinguish three cases:
(1) $m_i=s_i$; (2) $-s_i<m_i<s_i$ and (3) $m_i=-s_i$,
where $s_i\ne0$, with
the respective eigenvalues
\begin{subequations}\label{eq:SigmaSquareEigenvalue_i}
\begin{equation}\label{eq:SigmaSquareEigenvalue_mi=si}
\left( a^{s_i,s_i}_i\right)^2 s^2_i+a^{s_i,s_i}_i a^{s_i,s_i-1}_i s_i,
\end{equation}
\begin{align}\label{eq:SigmaSquareEigenvalue_mi}
& \tfrac12 a^{s_i,m_i}_i \left[a^{s_i,m_i+1}_i + 
a^{s_i,m_i-1}_i\right]s_i(s_i+1)
+\left( a^{s_i,m_i}_i\right)^2 m^2_i
\nonumber \\
& - \tfrac12 a^{s_i,m_i}_i m_i \left[a^{s_i,m_i+1}_i(m_i+1) 
+ a^{s_i,m_i-1}_i (m_i-1)\right],
\end{align}
\begin{equation}\label{eq:SigmaSquareEigenvalue_mi=-si}
\left( a^{s_i,-s_i}_i\right)^2 s^2_i+a^{s_i,-s_i}_i a^{s_i,1-s_i}_i s_i.
\end{equation}
\end{subequations}

On the other hand when the spins of $i$th and $j$th magnetic ions are
coupled, with total spin operator
$\hat{\mathbf{s}}_{ij}=\hat{\mathbf{s}}_{i}+\hat{\mathbf{s}}_{j}$,
the relation \eqref{eq:Sigma_i} enters a more general and complex expression.
To explore the properties of the coupled spins one has to
work with the total $\sigma$-operator $\hat{\boldsymbol{\sigma}}_{ij}$.
Its $z$ component and square are completely determined in the basis of the spin operator $\hat{\mathbf{s}}^2_{ij}$.
Similar to Eq. \eqref{eq:Sigma_i} for all $i\ne j$ and 
$\alpha \in \{x,y,z\}$, we have
\begin{equation}\label{eq:Sigma_ij}
\hat{\sigma}^\alpha_{ij} \lvert \ldots ,s_{ij},m_{ij}, \ldots \rangle
=
a^{s_{ij},m_{ij}}_{ij} \hat{s}^{\alpha_{\phantom{j}}}_{ij} 
\lvert \ldots ,s_{ij},m_{ij}, \ldots \rangle,
\end{equation}
where $a^{s_{ij},m_{ij}}_{ij} \in \mathbb{R}$. 
The corresponding rising and lowering operators obey
\begin{equation}\label{eq:Sigma_LR_ij}
\hat{\sigma}^{\pm}_{ij} \lvert \ldots ,s_{ij},m_{ij}, \ldots \rangle
=
a^{s_{ij},m_{ij}}_{ij} \hat{s}^{{\pm}_{\phantom{i}}}_{ij} 
\lvert \ldots ,s_{ij},m_{ij}, \ldots \rangle.
\end{equation}
The eigenvalues of $\hat{\boldsymbol{\sigma}}^2_{ij}$ depend on $m_{ij}$.
Therefore having in mind the following three cases $m_{ij}=s_{ij}$,
$-s_{ij}<m_{ij}<s_{ij}$ and $m_{ij}=-s_{ij}$, where $s_{ij}\ne0$ the eigenvalues read
\begin{subequations}\label{eq:SigmaSquareEigenvalue_ij}
\begin{equation}\label{eq:SigmaSquareEigenvalue_mij=sij}
\left( a^{s_{ij},s_{ij}}_{ij}\right)^2 s^2_{ij}+a^{s_{ij},s_{ij}}_{ij} a^{s_{ij},s_{ij}-1}_{ij} s_{ij},
\end{equation}
\begin{align}\label{eq:SigmaSquareEigenvalue_mij}
& \tfrac12 a^{s_{ij},m_{ij}}_{ij} \left[a^{s_{ij},m_{ij}+1}_{ij} + 
a^{s_{ij},m_{ij}-1}_{ij}\right]s_{ij}(s_{ij}+1)+
\left( a^{s_{ij},m_{ij}}_{ij}\right)^2 m^2_{ij}
\nonumber \\
& - \tfrac12 a^{s_{ij},m_{ij}}_{ij} m_{ij} 
\left[a^{s_{ij},m_{ij}+1}_{ij}(m_{ij}+1) 
+ a^{s_{ij},m_{ij}-1}_{ij} (m_{ij}-1)\right],
\end{align}
\begin{equation}\label{eq:SigmaSquareEigenvalue_mij=-sij}
\left( a^{s_{ij},-s_{ij}}_{ij}\right)^2 s^2_{ij}+a^{s_{ij},-s_{ij}}_{ij} a^{s_{ij},1-s_{ij}}_{ij} s_{ij}.
\end{equation}
\end{subequations}
The corresponding $\sigma$-operators share a single
coefficient and for $i \ne j$ and $\alpha \in \{x,y,z\}$, 
we have
\begin{equation}\label{eq:Sigma_k}
\hat{\sigma}^{\alpha}_i \lvert \ldots ,s_{ij},m_{ij}, \ldots \rangle
	=
a^{s_{ij},m_{ij}}_{ij} \hat{s}^{\alpha_{\phantom{j}}}_i 
\lvert \ldots ,s_{ij},m_{ij}, \ldots \rangle.
\end{equation}

We further assume that
the $\sigma$-operators preserve the corresponding
spin magnetic moment and for a noncoupled spin
obey the following constraints
\begin{subequations}\label{eq:Constraint_i}
\begin{equation}\label{eq:SigmaZ_i}
\hat{\sigma}^z_i \lvert \ldots ,s_i,m_i, \ldots \rangle
=
m_i \lvert \ldots ,s_i,m_i, \ldots \rangle,
\end{equation}
\begin{equation}\label{eq:SigmaSquare_i}
\hat{\boldsymbol{\sigma}}^2_i \lvert \ldots ,s_i,m_i, \ldots \rangle
=
s_i(s_i+1) \lvert \ldots ,s_i,m_i, \ldots \rangle.
\end{equation}
\end{subequations}
Similarly, when the $i$th and $j$th spins are 
coupled, for all $i \ne j$ we have
\begin{subequations}\label{eq:Constraint_ij}
\begin{equation}\label{eq:SigmaZ_ij}
\hat{\sigma}^z_{ij}
\lvert\ldots,s_{ij},m_{ij},\ldots \rangle =
m_{ij} \lvert\ldots,s_{ij},m_{ij},\ldots \rangle,
\end{equation}
\begin{equation}\label{eq:SigmaSquare_ij}
\hat{\boldsymbol{\sigma}}^2_{ij} 
\lvert\ldots,s_{ij},m_{ij},\ldots \rangle =
s_{ij}(s_{ij}+1) 
\lvert\ldots,s_{ij},m_{ij},\ldots \rangle.
\end{equation}
\end{subequations}
%Note that Eqs. \eqref{eq:SigmaSquare_ij} are regarded only with respect to
%eq. \eqref{eq:Sigma_k} and

Taking into account \eqref{eq:Constraint_i} together with
expressions \eqref{eq:SigmaSquareEigenvalue_i}
for all $i$ we have 
\begin{equation}\label{eq:ValueOf-ai}
\begin{array}{c}
a^{s_i,m_i\pm1}_i=a^{s_i,m_i}_i=1 \quad \forall \ m_i \ne0,
\\[0.1cm] 
a^{s_i,m_i\pm1}_i=a^{s_i,0}_i=\pm1.
\end{array}
\end{equation}
Further, according to constraints \eqref{eq:Constraint_ij} and Eqs. 
\eqref{eq:SigmaSquareEigenvalue_ij} we distinguish three cases:
%\begin{itemize}
%\item 

\textbf{(1)} $s_{ij} \ne 0$, $m_{ij} \ne 0$:
Then
\[
a^{s_{ij},m_{ij}\pm1}_{ij}=a^{s_{ij},m_{ij}}_{ij}=1.
\]
As a result the transformations of eigenvectors
via the $\sigma$-operator coincide with those defined by 
its corresponding spin operator.
Therefore, all couplings will be constants and the Hamiltonian
\eqref{eq:AddHamiltonian} will capture the same features as its
Heisenberg parent.

\textbf{(2)} $s_{ij} \ne 0$ and $m_{ij}=0$:
The corresponding coefficient cannot be determined from Eq.
\eqref{eq:SigmaZ_ij} and from Eqs. \eqref{eq:SigmaSquareEigenvalue_mij} and
\eqref{eq:SigmaSquare_ij} one obtains
\begin{equation}\label{eq:as0_ij}
a^{s_{ij},m_{ij}\pm1}_{ij}=a^{s_{ij},0}_{ij}=\pm1.
\end{equation}
We would like to point out that the ``minus''
sign is an intrinsic feature of the sigma operators
and is not related to the effectively 
accounted for spatial part of the wave function.

\textbf{(3)} $s_{ij} = 0$: The associated parameter
remains unconstrained and there exist a
set of coefficients $c^{n}_{ij}\in\mathbb{R}$ $\forall n\in\mathbb{N}$, 
such that
\begin{equation}\label{eq:a00_ij}
a^{0,0}_{ij} \in \{c^{n}_{ij}\}_{n\in\mathbb{N}}.
\end{equation} 
The values of $c^{n}_{ij}$ are indirect measures 
for the field strength along each possible exchange pathway 
and therefore the changes in the energy of exchange.
Depending on the type of exchange these effective 
coefficients are functions of the Coloumb, hopping and 
exchange integrals.
Thus, one can expect the emergence of bands
in the energy spectrum, associated with the 
existence of more then one exchange
pathway between
%transition metal
magnetic ions leading to a
broadened excitation width of the transition energy. 
Thereby, for a linear cluster with only one bonding anion between
magnetic cations one would obtain the limit
$|c^{n}_{ij}-c^{k}_{ij}| \to 0, \ \forall \ n \ne k$, 
where $c^{n}_{ij}\to1$. Accordingly,
the changes in the exchange field could be considered as negligible pointing 
to sharpened peaks in the magnetic 	
spectrum.
On the other hand, the inequality $|c^{n}_{ij}-c^{k}_{ij}| > 0$ for all 
$n \ne k$, would have to be considered as a sign 
for the presence of exchange paths of different energy and therefore of increased
excitation width in energy.
As an example, if an exchange bridge has a complex chemical structure, 
then one may expect
that the exchange path through which the electrons hop and 
are exchanged is not unique. 
Hence, the existence of $n$ different paths can be accounted
for by Hamiltonian \eqref{eq:AddHamiltonian}, where
according to \eqref{eq:a00_ij}, the transition energy 
$E_{ij}$ corresponding to the exchange of electrons between the $i$th and 
$j$th ions is written as
\begin{equation}\label{eq:TranEnergy_a1}
E_{ij}(c^{n}_{ij})=\tfrac12 J_{ij} (1+3c^{n}_{ij}),
\qquad
a^{1,0}_{ij}=1
\end{equation}
and
\begin{equation}\label{eq:TranEnergy_a-1}
E_{ij}(c^{n}_{ij})=\tfrac12 J_{ij} (3c^{n}_{ij}-1),
\qquad
a^{1,0}_{ij}=-1.
\end{equation}
Thus, the set of values $E_{ij}(c^{n}_{ij})$ will correspond to a broadened
peaks in the magnetic spectrum. Since $E_{ij}(c^{n}_{ij})=2J_{c^{n}_{ij}}$,
where $J_{c^{n}_{ij}}$ is the $n$-th value of the exchange coupling
from \eqref{eq:TranEnergy_a1} and \eqref{eq:TranEnergy_a-1} we respectively obtain
\begin{equation}\label{eq:cij_a1}
c^{n}_{ij}= \frac{4}{3}\frac{J_{c^{n}_{ij}}}{J_{ij}}-\frac13,  
\qquad
J_{1}=J_{ij}, \ \
a^{1,0}_{ij}=1
\end{equation}
and
\begin{equation}\label{eq:cij_a-1}
c^{n}_{ij}= \frac{4}{3}\frac{J_{c^{n}_{ij}}}{J_{ij}}+\frac13,  
\qquad
J_{5/3}=J_{ij}, \ \
a^{1,0}_{ij}=-1.
\end{equation}
As we will see later this approach 
allows one to explain the experimentally observed splitting and 
broadness of magnetic spectra in the molecular magnet Ni$_4$Mo$_{12}$.
Furthermore, if $J_{ij}$ is restricted to nearest-neighbors,
$c^{n}_{ij}$ can be used to compute the amount of energy required to observe an exchange with 
next-nearest neighbor ions. Thus only one coupling
parameter would be necessary within the present formalism.
In such case equations \eqref{eq:cij_a1} and 
\eqref{eq:cij_a-1} read
\begin{equation}\label{eq:NNcij_a1}
c^{n}_{i,j+1}= \frac{4}{3}\frac{J_{c^{n}_{i,j+1}}}{J_{ij}}-\frac13,  
\qquad
a^{1,0}_{ij}=1
\end{equation}
and
\begin{equation}\label{eq:NNcij_a-1}
c^{n}_{i,j+1}= \frac{4}{3}\frac{J_{c^{n}_{i,j+1}}}{J_{ij}}+\frac13,  
\qquad
a^{1,0}_{ij}=-1,
\end{equation}
respectively. The couplings $J_{c^{n}_{i,j+1}}$
will represent the exchange constant between the next-nearest neighbors.
This important feature will be illustrated by determining the INS spectrum \cite{matsuda_magnetic_2005} 
for the trimeric compound $\mathrm{Pb_3Cu_3(PO_4)_4}$.

Therefore a remarkable feature of the present approach is that 
\textit{when the spin quantum number of coupled
spins vanishes or we have at hand singlet bonds, then
the relevant coefficients might be represented as either discrete 
or continuous quantities}.
With the Hamiltonian \eqref{eq:AddHamiltonian} the eigenvalues 
of all eigenstates associated to singlet bonds will be unique.

\begin{figure*}[ht!]
\centering
	\def\s{0.5} % define scale
	\def\a{2} 	% define arrow width	
	\def\l{1.2} % define line width
	\begin{tikzpicture}[scale=0.85]
\begin{scope}[xshift=0cm]
	% draw vertical axis
	\draw[->, very thick] (-0.2,-3)--(-0.2,3);

	% energy units
	\node[rotate=90]() at (-0.5,1){Energy [meV]};

	% Ca3Cu3(PO4)4 levels
	\draw [black, line width=\l pt,blue, domain=0:3] plot (\x, -\s*4.725); %E0
	\draw [black, line width=\l pt, domain=0:3] plot (\x, \s*4.725); %E3
	\draw [black, line width=\l pt, domain=0:1.1] plot (\x, \s*4.725*0.2935); %{E1,E2}

	\draw [black, line width=\l pt,red, dashed] (1.1,\s*4.725*0.2935)--(1.4,\s*4.725*0.27)--(3,\s*4.725*0.27); %E1?
	\draw [black, line width=\l pt] (1.1,\s*4.725*0.2935)--(1.4,\s*4.725*0.317)--(3,\s*4.725*0.317); %E2
	
	% Ca3Cu3(PO4)4 transitions
	\draw[->, >=stealth,line width=\a pt, blue] (2,-\s*4.725) to (2,\s*4.725*0.317); %E20
	\draw[->, >=stealth,line width=\a pt, blue] (2.5,-\s*4.725) to (2.5,\s*4.725); %E30
	
	% Ca3Cu3(PO4)4 energy values
	\node[]() at (3.6,-\s*4.725){{\scriptsize $-7.09$}}; %E0=
	\node[]() at (3.6,\s*4.725){{\scriptsize $7.09$}}; %E3=
	\node[]() at (3.6,\s*4.725*0.317+0.15){{\scriptsize $2.25$}}; %E2=

	% spin quantum numbers
	\node[]() at (0.8,-\s*4.725+0.25){{\scriptsize $s_{13}=1$, $s=\frac{1}{2}$}}; %E0
	\node[]() at (0.8,\s*4.725*0.317+0.35){{\scriptsize $s_{13}=0$, $s=\frac{1}{2}$}}; %E1
	\node[]() at (0.8,\s*4.725+0.25){{\scriptsize $s_{13}=1$, $s=\frac{3}{2}$}}; %E3=

	% leybal Ca3Cu3(PO4)4
	\node[]() at (1.8,3.7){{\small $\mathrm{Ca_3Cu_3(PO_4)_4}$}};
\end{scope}
\begin{scope}[xshift=5cm]
	% Sr3Cu3(PO4)4 levels 
	\draw [black, line width=\l pt,blue, domain=0:3] plot (\x, -\s*5.021); %E0
	\draw [black, line width=\l pt, domain=0:3] plot (\x, \s*5.021); %E3
	\draw [black, line width=\l pt, domain=0:1.1] plot (\x, \s*5.021*0.2965); %{E1,E2}
	
	\draw [black, line width=\l pt,red, dashed] (1.1,\s*5.021*0.2965)--(1.4,\s*5.021*0.272)--(3,\s*5.021*0.272); %E1?
	\draw [black, line width=\l pt] (1.1,\s*5.021*0.2945)--(1.4,\s*5.021*0.319)--(3,\s*5.021*0.319); %E2
	
	% Sr3Cu3(PO4)4 transitions
	\draw[->, >=stealth, line width=\a pt, blue] (2,-\s*5.021) to (2,\s*5.021*0.319); %E20
	\draw[->, >=stealth, line width=\a pt, blue] (2.5,-\s*5.021) to (2.5,\s*5.021); %E30

	% spin quantum numbers
	\node[]() at (0.8,-\s*5.021+0.25){{\scriptsize $s_{13}=1$, $s=\frac{1}{2}$}}; %E0
	\node[]() at (0.8,\s*5.021*0.317+0.35){{\scriptsize $s_{13}=0$, $s=\frac{1}{2}$}}; %E1
	\node[]() at (0.8,\s*5.021+0.25){{\scriptsize $s_{13}=1$, $s=\frac{3}{2}$}}; %E3=	
	
	% Sr3Cu3(PO4)4 energy values
	\node[]() at (3.6,-\s*5.021){{\scriptsize $-7.53$}}; %E0=
	\node[]() at (3.6,\s*5.021){{\scriptsize $7.53$}}; %E3=
	\node[]() at (3.6,\s*5.021*0.319+0.15){{\scriptsize $2.4$}}; %E2=
	
	% leybal Sr3Cu3(PO4)4
	\node[]() at (1.8,3.7){{\small $\mathrm{Sr_3Cu_3(PO_4)_4}$}};
\end{scope}
\begin{scope}[xshift=10cm]
	% Pb3Cu3(PO4)4 levels 
	\draw [black, line width=\l pt, blue, domain=0:3] plot (\x, -\s*4.564); %E0
	
	\draw [black, line width=\l pt, domain=0:3] plot (\x, \s*4.564); %E3
	
	\draw [black, line width=\l pt, domain=0:1.1] plot (\x, \s*4.564*0.295); %{E1,E2}
	
	\draw [black, line width=\l pt,red] (1.1,\s*4.564*0.295)--(1.4,\s*4.564*0.274)--(3,\s*4.564*0.274); %E1
	\draw [black, line width=\l pt] (1.1,\s*4.564*0.295)--(1.4,\s*4.564*0.325)--(3,\s*4.564*0.325); %E2
	
	% Pb3Cu3(PO4)4 transitions
	\draw[->, >=stealth, line width=\a pt, blue] (1.8,-\s*4.564) to (1.8,\s*4.564*0.325); %E20
	
	\draw[->, >=stealth, line width=\a pt, blue] (2.6,-\s*4.564) to (2.6,\s*4.564); %E30
	
	\draw[->, >=stealth, line width=\a pt, red] (2.2,\s*4.564*0.274) to (2.2,\s*4.564); %E31
	
	% Pb3Cu3(PO4)4 energy values
	\node[]() at (3.6,-\s*4.564){{\scriptsize $-6.85$}}; %E0=
	\node[]() at (3.6,\s*4.564){{\scriptsize $6.85$}}; %E3=
	\node[]() at (3.6,\s*4.564*0.315+0.15){{\scriptsize $2.16$}}; %E2=
	\node[]() at (3.6,\s*4.564*0.284-0.15){{\scriptsize $1.95$}}; %E1=

	% spin quantum numbers
	\node[]() at (0.8,-\s*4.564+0.25){{\scriptsize $s_{13}=1$, $s=\frac{1}{2}$}}; %E0
	\node[]() at (0.8,\s*4.564*0.317+0.35){{\scriptsize $s_{13}=0$, $s=\frac{1}{2}$}}; %E1
	\node[]() at (0.8,\s*4.564+0.25){{\scriptsize $s_{13}=1$, $s=\frac{3}{2}$}}; %E3=
	
	% leybal Pb3Cu3(PO4)4
	\node[]() at (1.8,3.7){{\small $\mathrm{Pb_3Cu_3(PO_4)_4}$}};
\end{scope}
\begin{scope}[xshift=16cm]
% leybaling energy levels

\node[]() at (0.2,\s*4.55){{\scriptsize $\big\{a^{1,m_{13}}_{13},a^{1/2,m_{2}}_2\big\}=\{1,1\}$}}; % quartet

\node[]() at (0.25,\s*4.7*0.4){{\scriptsize 
$\big\{a^{0,0}_{13},a^{1/2,\pm 1/2}_2\big\}=\{c^2_{13},1\}$}}; %doublet

\node[]() at (0.25,\s*4.7*0.2){{\scriptsize 
$\big\{a^{0,0}_{13},a^{1/2,\pm 1/2}_2\big\}=\{c^1_{13},1\}$}}; %doublet

\node[]() at (0.2,-\s*4.7){{\scriptsize $\big\{a^{1,m_{13}}_{13},a^{1/2,m_{2}}_2\big\}=\{1,1\}$}}; %ground state(doublet)
\end{scope}
\end{tikzpicture}
\caption{Energy level structure of the compounds
$\mathrm{A_3Cu_3(PO_4)_4}$ (A = Ca, Sr, Pb). The blue arrows
show the ground state transitions, and red arrow stands for the excited
transition. The energy levels corresponding to the ground state are
designated by blue lines.
The initial energy level of the excited transition is depicted by
a red line, while by analogy to $\mathrm{Pb_3Cu_3(PO_4)_4}$ the dashed red lines
stand for a presumed second sub level of the excited doublet level.
}
\label{fig:CopperEnergyStructure}
\end{figure*}
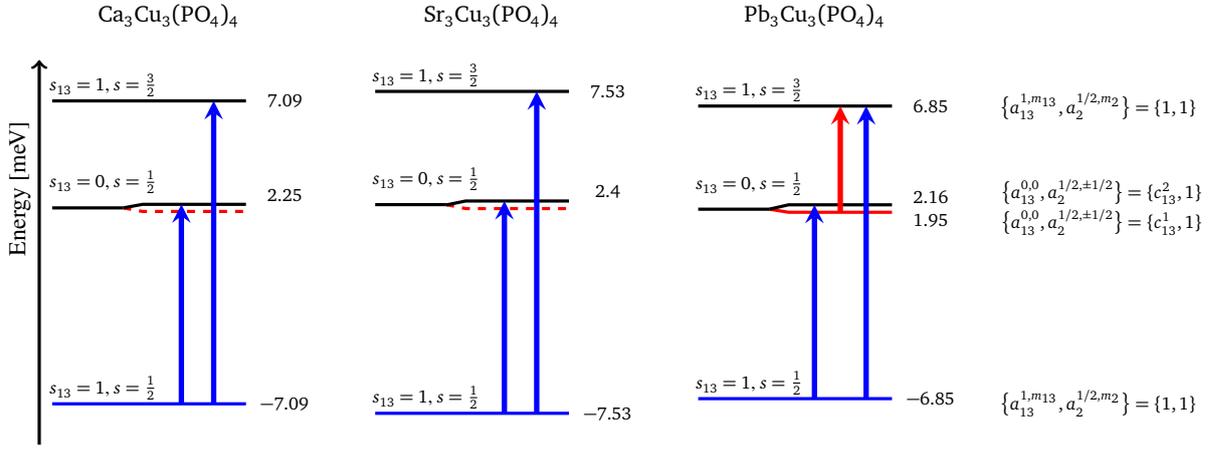

\section{A$_3$C\lowercase{u}$_3$(PO$_4$)$_4$ (A =
	C\lowercase{a}, S\lowercase{r} \lowercase{and} P\lowercase{b})} \label{sec:copper}

\subsection{The Hamiltonian}
The magnetic compounds A$_3$Cu$_3$(PO$_4$)$_4$ (A = Ca, Sr, Pb) are
convenient spin trimer systems, with spin-$\tfrac12$ Cu$^{2+}$, for
testing the Hamiltonian \eqref{eq:AddHamiltonian} and studying
the fundamental nature of antiferromagnetism.
FIG. \ref{fig:Copper} (a) shows a small fragment of the
copper ions structure with the exchange pathways relevant to oxygen
atoms arrangements, where Cu2 ion is surrounded by four oxygen atoms
on a plane, while Cu1 and Cu3 ions are surrounded by five oxygen atoms
constructing distorted square pyramid. For brevity the other elements
are not shown and only two oxygen atoms along the intratrimer
Cu1--O1--Cu2 and intertrimer Cu2--O2--Cu4 pathways are labelled. In
general, the exchange processes appear to be more complex and
depend on the global structure of the compounds
\cite{drillon_1d_1993}. Besides the
superexchange interactions are sensitive \cite{matsuda_magnetic_2005} to the angle between
$\mathrm{Cu^{2+}}$ bonds and their lengths suggesting that the intertrimer
Cu2--Cu4 interaction is much smaller than the
intratrimer ones \textit{i.e.} Cu1--Cu2 and Cu3--Cu2. Thus, the intertrimer
exchange can be neglected and the $\mathrm{Cu^{2+}}$ sub-lattice is
considered as a one-dimensional array of isolated
spin trimers FIG. \ref{fig:Copper} (b).

Applying the formalism of Section \ref{sec:spinmodel} 
by considering equations \eqref{eq:NNcij_a1}, \eqref{eq:NNcij_a-1}
and taking into account that Cu1-Cu2 and Cu2-Cu3 are bonded by a single 
oxygen ion, we set $J_{ij}\to J_{12}=J$ and
perform a study of the magnetic excitations.
Owing to the trimer symmetry,
Hamiltonian \eqref{eq:AddHamiltonian} transforms into
\begin{equation}\label{eq:CopperHamilton}
\hat{\mathcal{H}} = J \left( \hat{\boldsymbol{\sigma}}_{13}\cdot\hat{\mathbf{s}}_2 + 
\hat{\boldsymbol{\sigma}}_2\cdot\hat{\mathbf{s}}_{13} + 
\hat{\boldsymbol{\sigma}}_1\cdot\hat{\mathbf{s}}_3+
\hat{\boldsymbol{\sigma}}_3\cdot\hat{\mathbf{s}}_1\right). 
\end{equation}
With respect to Eq. \eqref{eq:ValueOf-ai} the
total spin eigenstates are denoted by $\rvert s_{13},s,m\rangle$.
Hence in contrast to the eigenvalues of \eqref{eq:MainHamiltonian} obtained in Refs.
\cite{furrer_magnetic_2010,machens_even-odd_2013,podlesnyak_magnetic_2007},
the eigenvalues of Hamiltonian \eqref{eq:CopperHamilton} have an additional
parameter that can be tuned to identify the
energy of the experimentally observed third (excited) transition
\cite{matsuda_magnetic_2005}.

\subsection{Energy levels}
According to \eqref{eq:ValueOf-ai} we 
have $a^{s_2,m_2}_2=1$ for all energy levels.
When the spin cluster is characterized by triplet states
$\big\lvert 1, {\tfrac{1}{2}}, \pm\tfrac{1}{2} \big\rangle$,
for all $m_{ij}$ we have $a^{1,m_{ij}}_{13}=1$. Thus, taking into account
\eqref{eq:CopperHamilton} we obtain the ground state energy
\begin{equation}\label{eq:CopperEnergyGS}
E^{\pm 1/2}_{1,1/2} = -\tfrac32 J.
\end{equation}
The second pair of doublet states is
associated with the first excited energy level, see FIG.
\ref{fig:CopperEnergyStructure}. 
The edged spins of the isolated trimer are coupled in a singlet,
with corresponding state $\big\lvert 0, {\tfrac{1}{2}}, \pm\tfrac{1}{2} \big\rangle$, 
\textit{i.e.} $m_{13} = 0$, $s_{13} = 0$.
Now, using
\eqref{eq:CopperHamilton} we end up with
\begin{equation}\label{eq:CopperEnergyFirstEx}
E^{\pm 1/2}_{0,1/2} = - \tfrac32J a^{0,0}_{13}.
\end{equation}
To fully characterize the experimentally observed transitions 
for $\mathrm{Pb_3Cu_3(PO_4)_4}$ one requires at least three excited 
energy levels. Bearing in mind that the quartet level is
four-fold degenerate, we
deduce that the corresponding coefficient may take only two values 
$a^{0,0}_{13}\in\{c^1_{13},c^2_{13}\}$. 
Further, the observed excitations spectra
\cite{matsuda_magnetic_2005} are not broadened signalling that
$|c^1_{13}-c^2_{13}| \approx 0$.
Therefore taking into account \eqref{eq:CopperEnergyFirstEx} we get
\[
E^{\pm 1/2}_{0,1/2} \in \left\lbrace -\tfrac32 J c^1_{13},-\tfrac32 J 
c^2_{13}\right\rbrace .
\] 	
Furthermore, in the quartet eigenstate with all spins pointing to the
same direction,
$m_{13}=\pm 1$ and $m_2 = \pm \tfrac12$. Thus, 
the trimer is in the state $\big\lvert 1, {\tfrac{3}{2}}, \pm\tfrac{3}{2} \big\rangle$ 
and the energy reads
\[
E^{\pm 3/2}_{1,3/2} = \tfrac12 J\left( 1+
a^{1,\pm 1}_{13} \right) + \tfrac12 Ja^{1,\pm 1}_{13} 
= \tfrac32 J.
\]
For the remaining two quartet eigenstates with $m=\pm\tfrac12$,
for all $m_{ij}$ we have $a^{1,m_{ij}}_{13}=1$ thus
\[
E^{\pm 1/2}_{1,3/2}=\tfrac32 J.
\]
Whence, the energy sequence consists of four levels.
Henceforth we denote these levels as follow 
\begin{equation}\label{eq:CopperEnergyNotation}
E_0 = -\tfrac32 J, \quad E_1 = -\tfrac32 J c^1_{13},
\quad E_2 = -\tfrac32 J c^2_{13}, \quad E_3 = \tfrac32 J.
\end{equation}

\subsection{Scattering intensities}	
The corresponding
selection rules are $\Delta s_{13} = 0, \pm 1$, $\Delta s = 0, \pm 1$
and $\Delta m = 0, \pm 1$.
Calculating the scattering functions in
\eqref{eq:ScatterFunctions}
with $\lvert n \rangle \equiv \lvert s_{13},s,m\rangle$,
for transitions between the energy levels, we get $S^{\alpha
\beta}(\mathbf{q},\omega_{n'n})+S^{\beta
\alpha}(\mathbf{q},\omega_{n'n})=0$, and $S^{\alpha
\alpha}(\mathbf{q},\omega_{n'n})=S^{\beta
\beta}(\mathbf{q},\omega_{n'n})$ for all $\alpha, \beta$ and 
$n,n' = 0,1,2,3$. 
Moreover, taking into account to the cluster structure,
we have $\sum_{\alpha}\Theta^{\alpha \alpha} = 2$.
Note that due to the degeneracy of the energy spectrum with
respect to $m$, for each value of $s$ and $s_{13}$ the summation over
$n$ and $n'$ in \eqref{eq:ScatterFunctions} corresponds to a
summation over all possible values of the total magnetic quantum number. The
analysis of the intensities, taking into account the experimental data,
allows us to determine the observed first magnetic excitation
corresponding to the transition between 
the ground state $\big\lvert 1, {\tfrac{1}{2}}, \pm\tfrac{1}{2} \big\rangle$
and the first excited states $\big\lvert 0, {\tfrac{1}{2}},
\pm\tfrac{1}{2} \big\rangle$
with scattering functions
\[
S^{\alpha \alpha}(\mathbf{q},\omega_{20}) =\tfrac{1}{3} 
[1-\cos(2\mathbf{q}\cdot\mathbf{r})]p_0,
\]
where $\mathbf{r}$ is the vector of the average distance between
neighboring ions with
$\mathbf{r}_{31}=2\mathbf{r}$. The rotational degeneracy of the
quartet energy level is four--fold and hence the second ground
state excitation refer to transitions from the 
doublet 
$\big\lvert 1, {\tfrac{1}{2}}, \pm\tfrac{1}{2} \big\rangle$
to the quartet states $\big\lvert 1, {\tfrac{3}{2}}, m \big\rangle$,
where $m=\pm\tfrac12, \pm\tfrac32$. Accordingly, we get
\[
S^{\alpha \alpha}(\mathbf{q},\omega_{30}) = \tfrac{2}{9} 
[3 + \cos(2\mathbf{q}\cdot\mathbf{r}) - 4\cos(\mathbf{q}\cdot\mathbf{r})]p_0.
\]
The excited peak is indicated by the transitions between the doublet 
$\big\lvert 0, {\tfrac{1}{2}}, \pm\tfrac{1}{2} \big\rangle$
and the quartet eigenstates 
$\big\lvert 1, {\tfrac{3}{2}}, m \big\rangle$.
The corresponding scattering functions are
\[
S^{\alpha \alpha}(\mathbf{q},\omega_{31}) = \tfrac{2}{3}[1 - 
\cos(2\mathbf{q}\cdot\mathbf{r})]p_1.
\]
Therefore, according to Eqs. \eqref{eq:ScateringIntensities}
we estimate the relevant intensities
\begin{equation}\label{eq:CopperIntegIntensities}
\begin{array}{l}
\displaystyle
I_{20} \propto  \gamma_{20}
 \left[ 1-\frac{\sin(2\mathrm{q}\mathrm{r})}{2\mathrm{q}\mathrm{r}}\right] 
 F^2(\mathrm{q}),
\\ [0.3cm]
\displaystyle
I_{30} \propto \gamma_{30} 
 \left[ 1 + \frac{\sin(2\mathrm{q}\mathrm{r})}{6\mathrm{q}\mathrm{r}} - 4\frac{\sin(\mathrm{q}\mathrm{r})}{3\mathrm{q}\mathrm{r}}\right] 
 F^2(\mathrm{q}),
\\[0.3cm]
\displaystyle
I_{31} \propto  \gamma_{31} 
 \left[ 1-\frac{\sin(2\mathrm{q}\mathrm{r})}{2\mathrm{q}\mathrm{r}}\right]
  F^2(\mathrm{q}),
\end{array}
\end{equation}
where
\[
\gamma_{20} = \tfrac23 p_0, \ \gamma_{30} = \tfrac{12}9 p_0, \ 
\gamma_{31} = \tfrac43 p_1.
\]
Now, substituting the first radial wave function $R_{10}$ 
and the first spherical Bessel function $J_0$ in Eq.
\eqref{eq:MagFormFactor}, for dications Cu$^{2+}$, we have 
\begin{equation}\label{eq:MagFormFactorIntegrate}
F(\mathrm{q}) = \frac{256}{(16+\mathrm{q}^2 \mathrm{r}^2_{\mathrm{o}})^2},
\end{equation}
where $\mathrm{r}_{\mathrm{o}} = 0.529 \, \mathrm{\mathring{A}}$ is the Bohr radius.
\begin{figure}[!h]
\centering
\includegraphics[scale=1]{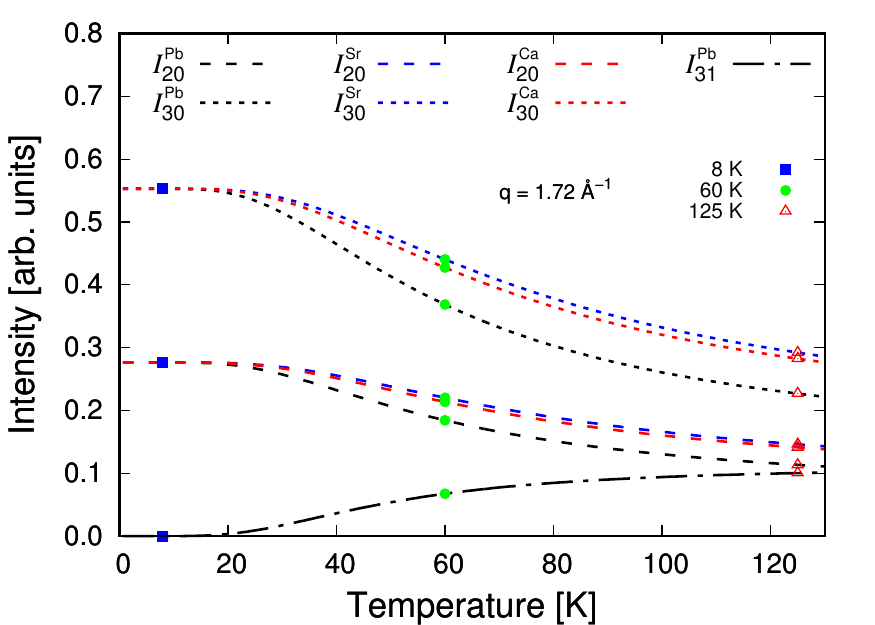}
\caption{
Scattering intensities $I^{\mathrm{A}}_{20}$ and $I^{\mathrm{A}}_{30}$,
with (A = Ca, Sr, Pb), of the ground state transitions as a function
of the temperature. 
The blue squares, the green
circles and red triangles correspond to the values of the intensities
given in TAB. \ref{tab:CopperIntensThreeTemp}.
\label{fig:CuIntensity}}
\end{figure}

\begin{figure}[h!]
\centering
\includegraphics[scale=1]{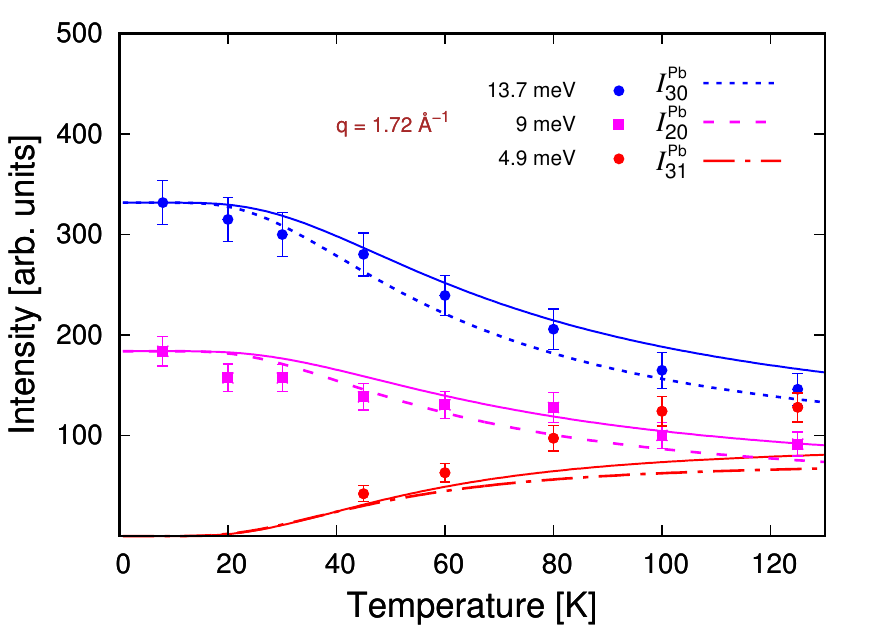}
\caption{Scattering intensities for the compound 
Pb$_3$Cu$_3$(PO$_4$)$_4$ as a function of the temperature,
along with experimental results from Ref. \cite{matsuda_magnetic_2005}.
The solid and dashed lines show the calculated intensities
for the
Heisenberg model and Hamiltonian
\eqref{eq:AddHamiltonian}, respectively.
\label{PbIntensity}}
\end{figure}

\subsection{Energy of the magnetic transitions}	
Denoting the energies of transitions between energy
levels by $E_{ij}$ we get
%and taking into account eq. [?] we get
\begin{equation}\label{eq:CopperGSTransitions}
E_{20} = \tfrac32 J \left(1 - c^2_{13}\right) ,
\quad
E_{30} = 3 J,
\quad
E_{31} = \tfrac32 J \left( 1 + c^1_{13}\right) .
\end{equation}
Neutron
scattering experiments performed on $\mathrm{Pb_3Cu_3(PO_4)_4}$ with
$T \geq 60$ K \cite{matsuda_magnetic_2005} shows the
presence of a third peak at about $4.9$ meV, which may be related
to the excited transition energy $E_{31}$. 
The values of $c^1_{13}$,
$c^2_{13}$ and $J$, according to INS experiments
\cite{matsuda_magnetic_2005} performed on polycrystalline samples
$\mathrm{A_3Cu_3(PO_4)_4}$ (A = Ca, Sr, Pb) are shown in TAB.
\ref{tab:Copper}. In addition, for the compound $\mathrm{Ca_3Cu_3(PO_4)_4}$
we have $c^2_{13} = -0.32(8)$ and $J \approx 4.741 \, \mathrm{meV}$
based on INS data at $T=1.5$ K
\cite{furrer_magnetic_2010,podlesnyak_magnetic_2007}.
\begin{table}[ht!]
\caption{
The values of the coupling constant and the quantities $c^1_{13}$, $c^2_{13}$
for $\mathrm{A_3Cu_3(PO_4)_4}$ (A = Ca, Sr, Pb)
obtained by taking into account the experimental data 
in Ref. \cite{matsuda_magnetic_2005}. 
\label{tab:Copper}
}
\begin{ruledtabular}%	\toprule[1.2pt]
\begin{tabular}{ccccccccc}
%	{} & [meV] & [meV] & [meV] & {} & {} & [meV] & [meV] & \\
 A & $E_{20}$ & $E_{30}$  
& $E_{31}$ & $c^1_{13}$ & $c^2_{13}$ & $J$ & $J_{c^2_{13}}$ & $J_{c^1_{13}}$ \\
\hline %		\midrule[0.5pt]	
Ca & 9.335 & 14.174 & $-$ & $-$ & -0.317 & 4.725 & 0.058 & --\\
Sr & 9.936 & 15.064 & $-$ & $-$ & -0.319 & 5.021 & 0.054 & --\\
Pb & 9.005 & 13.693 & 4.9 & -0.284 & -0.315 & 4.564 & 0.062 & 0.168\\ 
\end{tabular}
\end{ruledtabular}%		\bottomrule[1.2pt]
\end{table}

\begin{figure}[ht!]
\centering
\includegraphics[scale=1]{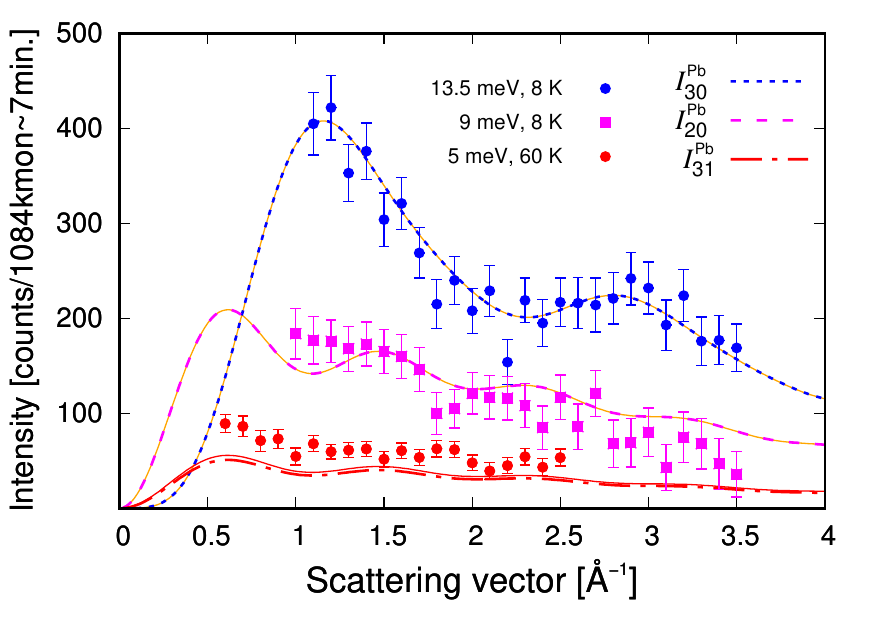}
\caption{
Calculated intensities in arbitrary units as a function of the scattering vector
for Pb$_3$C\lowercase{u}$_3$(PO$_4$)$_4$ along with the experimental
data of Ref. \cite{matsuda_magnetic_2005}. 
The dashed lines depict the intensities obtained from the 
Hamiltonian \eqref{eq:AddHamiltonian}. 
The solid red and orange lines correspond to the Heisenberg model.
$I^{\mathrm{Pb}}_{20}$ and $I^{\mathrm{Pb}}_{30}$ correspond to the ground state
transitions at $T=8$ K.
The intensity $I^{\mathrm{Pb}}_{31}$ stands for the excited transition
at $T=60$ K. Our results show that the theoretical curves for all compounds
coincide.
\label{fig:CuFFIntensity}
}
\end{figure}

The temperature dependence of the integrated scattering intensities for each
compound is shown on FIG. \ref{fig:CuIntensity} obtained with 
the form factor \eqref{eq:MagFormFactorIntegrate}. On
FIG. \ref{PbIntensity} we present the scattering intensities for 
$\mathrm{Pb_3Cu_3(PO_4)_4}$ computed with our Hamiltonian and
the Heisenberg model along with the
experimental data taken from Ref. \cite{matsuda_magnetic_2005}. 
Let us point out that
our results are in better agreement with their
experimental counterpart for $I_{20}^\mathrm{Pb}$ and $I_{30}^\mathrm{Pb}$, while for $I_{31}^\mathrm{Pb}$
we have a qualitative agreement. The averaged
magnitudes of the scattering vector $q$ and the distance $r$ between
neighboring ions are taken from Ref.
\cite{matsuda_magnetic_2005}, $q=1.72\ \mathrm{\mathring{A}}^{-1}$
and $r = 3.6 \ \mathrm{\mathring{A}}$. The explicit expressions of the
scattering intensities for each transition are
\begin{subequations}
\begin{align}\label{eq:CrossSectionsCopper}
I^{\mathrm{A}}_{20}(T) & \propto 0.5528 Z^{-1}_\mathrm{A} e^{-\frac{E^{\mathrm{A}}_0}{k_BT}},
\\
I^{\mathrm{A}}_{30}(T) & \propto 1.1057 Z^{-1}_\mathrm{A} e^{-\frac{E^{\mathrm{A}}_0}{k_BT}},
\\
I^{\mathrm{Pb}}_{31}(T) & \propto 1.1056 Z^{-1}_{\mathrm{Pb}} e^{-\frac{E^{\mathrm{Pb}}_1}{k_BT}},
\end{align} 
\end{subequations}
where A = Ca, Sr, Pb. As $T$ vanishes the scattering intensities
of first and second transitions from the ground state to the excited states are equal by about a
factor of 2, see TAB. \ref{tab:CopperIntensThreeTemp}.
For $T > 20$ K a third peak
sets in, but the evaluated intensity $I^{\mathrm{Pb}}_{31}$ remains
smaller than the experimentally observed one
\cite{matsuda_magnetic_2005}. In contrast to the functions
$I^{\mathrm{Pb}}_{30}$ and $I^{\mathrm{Pb}}_{20}$ the intensities of
the ground state transitions for A = Ca, Sr
decrease slowly with temperature. The predicted peak for
$\mathrm{Pb_3Cu_3(PO_4)_4}$ is in concert with
the experimental findings \cite{matsuda_magnetic_2005}. 
Unfortunately there are no experimental data confirming the presence of this third peak for
the compounds $\mathrm{Ca_3Cu_3(PO_4)_4}$ and $\mathrm{Sr_3Cu_3(PO_4)_4}$
and hence the energy level $E_1$ could not be included in the sequence of energy spectrum.
On FIG. \ref{fig:CopperEnergyStructure} the presumed energy levels $E^{\mathrm{Ca}}_1$ and $E^{\mathrm{Sr}}_1$ are illustrated with dashed red lines.
For all compounds the scattering intensities as a function of the
magnitude of the scattering vector are represented in FIG. \ref{fig:CuFFIntensity}.

\begin{table}[!ht]
\caption{
Calculated values of integrated scattering intensities $I^{\mathrm{A}}_{n'n}$
[arb. units] with A = Ca, Sr, Pb at temperatures 8, 60
and 125 K, depicted on FIG. \ref{fig:CuIntensity}
\label{tab:CopperIntensThreeTemp}
}
\begin{ruledtabular}%	\toprule[1.2pt]
\begin{tabular}{lccc}
$T$ [K]                &    8     &     60   &    125   \\
\hline
$I^{\mathrm{Ca}}_{20}$ & 0.276(4) & 0.213(7) & 0.141(2) \\
$I^{\mathrm{Ca}}_{30}$ & 0.552(8) & 0.427(4) & 0.282(5) \\
$I^{\mathrm{Sr}}_{20}$ & 0.276(4) & 0.220(2) & 0.146(1) \\
$I^{\mathrm{Sr}}_{30}$ & 0.552(8) & 0.440(5) & 0.292(2) \\
$I^{\mathrm{Pb}}_{20}$ & 0.276(4) & 0.184(3) & 0.113(4) \\
$I^{\mathrm{Pb}}_{30}$ & 0.552(8) & 0.368(6) & 0.226(8) \\
$I^{\mathrm{Pb}}_{31}$ & 0        & 0.067(3) & 0.100(3) \\	
\end{tabular}
\end{ruledtabular}%	\toprule[1.2pt]
\end{table}	

\begin{figure*}[ht!]
\begin{tikzpicture}%[scale=0.85]
\def\a{2}	% define arrow width
\def\l{1.2}	% define line width
%-------------------------Singlet-----------------------------------
\draw [blue, line width=\l pt, domain=0:8] plot (\x, 0); %E0
\node[]() at (0.5,0.2){{\footnotesize $s=0$}}; %s=0

%		\node[]() at (3.5,-0.25){{\scriptsize $\lvert 1, 1, 0,0 \rangle$}};%E0
%-------------------------Triplet----------------------------------
\draw [black, line width=\l pt, domain=0:5] plot (\x, 2); %E4
\draw [red, line width=\l pt, domain=5:8] plot (\x, 2); %E4	

\node[]() at (0.5,2.2){{\footnotesize $s=1$}};%s=1

%-sub level 1---{s12,s34,s}={0,1,1},{0,2,2}----------
\draw [black,line width=\l pt] (1.5,2)--(2,1.1)--(5,1.1);%sub1

	\node[]() at (3.85,0.85){{\scriptsize $\lvert 0, 1, 1,m \rangle, \lvert 0, 2, 2,0 \rangle$}};%sub1
	
\draw [black,dashed,line width=\l pt] (5,1.1)--(5.5,1.3)--(8,1.3);%E2
\draw [black,line width=\l pt] (5,1.1)--(5.5,0.85)--(8,0.85);%E1
%----------------------------------------------------

%sub level 2---{s12,s34,s}={1,0,1},{2,0,2}-----------
\draw [black,line width=\l pt] (1.5,2)--(2,1.75)--(8,1.75);%E3

	\node[]() at (3.85,1.5){{\scriptsize $\lvert 1, 0, 1,m \rangle, \lvert 2, 0, 2,0 \rangle$}};%E3
%---------------------------------------------------- 

%-----------set connection to E5, E6-------
\draw [black,line width=\l pt] (1.5,2)--(2,3.1);

%-----------set connection to E7----------
\draw [black,line width=\l pt] (1.5,2)--(2,3.75);		

%-------------------------Quartet----------------------------------
\draw [black, line width=\l pt, domain=0:8] plot (\x, 4); %E8
\node[]() at (0.5,4.2){{\footnotesize $s=2$}}; %s=2

%-sub level 1---{s12,s34,s}={0,1,1},{0,2,2}-----------
\draw [black,line width=\l pt] (1.5,4)--(2,3.1)--(5,3.1);

	\node[]() at (3.85,2.85){{\scriptsize $\lvert 0, 1, 1,0 \rangle, \lvert 0, 2, 2,m \rangle$}};%sub1

\draw [black, dashed,line width=\l pt] (5,3.1)--(5.5,3.3)--(8,3.3);%E6
\draw [black,line width=\l pt] (5,3.1)--(5.5,2.85)--(8,2.85);%E5
%----------------------------------------------------

%sub level 2---{s12,s34,s}={1,0,1},{2,0,2}-----------
\draw [black,line width=\l pt] (1.5,4)--(2,3.75)--(8,3.75);

	\node[]() at (3.85,3.5){{\scriptsize $\lvert 1, 0, 1,0 \rangle, \lvert 2, 0, 2,m \rangle$}};%sub2
%---------------------------------------------------- 
	
%-----------set connection to E3----------
\draw [black,line width=\l pt] (1.5,4)--(2,1.75);

%-----------set connection to E1,E2----------
\draw [black,line width=\l pt] (1.5,4)--(2,1.1);

%-------------------------Septet----------------------------------
\draw [black, line width=\l pt, domain=0:8] plot (\x, 6); %E9
\node[]() at (0.5,6.2){{\footnotesize $s=3$}}; %s=3
%-------------------------Nonet-----------------------------------
\draw [black, line width=\l pt, domain=0:8] plot (\x, 8); %E10
\node[]() at (0.5,8.2){{\footnotesize $s=4$}}; %s=4
	
%--------------------transitions----------------------------------
\draw[->, >=stealth, line width=\a pt, blue] (5.75,0) to (5.75,0.85); %E10
\draw[->, >=stealth, line width=\a pt, blue] (6.375,0) to (6.375,1.75); %E30
\draw[->, >=stealth, line width=\a pt, blue] (7,0) to (7,2.85); %E50
\draw[->, >=stealth, line width=\a pt, red] (7.5,2) to (7.5,3.3); %E64

\node[]() at (5.95,0.3){{\small I}}; %First
\node[]() at (6.575,0.3){{\small II}}; %Second
\node[]() at (7.25,0.3){{\small III}}; %Third
\node[]() at (7.8,2.3){{\small IV}}; %Fourt
%-------------------draw vertical axis----------------------------
\draw[very thick] (8.45,-0.2)--(8.45,0.42); % axis for E0 to E1
\draw[very thick] (8.45,0.5)--(8.45,2.31); % axis for E1 to E4
\draw[very thick] (8.45,2.39)--(8.45,5.02); % axis for E4 to E8
\draw[very thick] (8.45,5.1)--(8.45,6.72); % axis for E9 to E10
\draw[->, very thick] (8.45,6.8)--(8.45,8.5); % axis for E8
\node[rotate=0]() at (9.4,8.5){{\small [meV]}}; 
	
%-------------------draw intersections--------------------------
\node[rotate=0]() at (8.45,0.445){{$\approx$}}; 
\node[rotate=0]() at (8.45,2.345){{$\approx$}}; 
\node[rotate=0]() at (8.45,5.045){{$\approx$}}; 
\node[rotate=0]() at (8.45,6.745){{$\approx$}}; 
	
%--------------------energy values--------------------------------
\node[]() at (9.5,0){{\footnotesize $E_0 =-2.6\phantom{a}$}}; %E0=
\node[]() at (9.5,0.85){{\footnotesize $E_1=-2.2\phantom{a}$}}; %E1=
\node[]() at (9.5,1.3){{\footnotesize $E_2=-2.1\phantom{a}$}}; %E2=
\node[]() at (9.5,1.75){{\footnotesize $E_3=-2\phantom{.aa}$}}; %E3=
\node[]() at (9.5,2.1){{\footnotesize $E_4=-1.95$}}; %E4=
\node[]() at (9.5,2.85){{\footnotesize $E_5=-0.9\phantom{a}$}}; %E5=
\node[]() at (9.5,3.3){{\footnotesize $E_6=-0.8\phantom{a}$}}; %E6=
\node[]() at (9.5,3.75){{\footnotesize $E_7=-0.7\phantom{a}$}}; %E7=
\node[]() at (9.5,4.1){{\footnotesize $E_8=-0.65$}}; %E8=
\node[]() at (9.5,6){{\footnotesize $E_9=1.3\phantom{-a}$}}; %E9=
\node[]() at (9.5,8){{\footnotesize $E_{10}=3.9\phantom{-a}$}}; %E10=
	
%--------------------parameter values--------------------------------
\node[]() at (12.45,0) {{\scriptsize 
$\big\{ a^{1, m_{12}}_{12},a^{1,m_{34}}_{34}\big\} =\{1,1\}$}}; %E0=

\node[]() at (13,0.825) {{\scriptsize
$\big\{ a^{0,0}_{12},a^{1,m_{34}}_{34},a^{2,0}_{34}\big\} =\{1.1923,1,-1\}$}}; %E1=
	
\node[]() at (13,1.325) {{\scriptsize 
$\big\{ a^{0,0}_{12},a^{1,m_{34}}_{34},a^{2,0}_{34}\big\} =\{1.1153,1,-1\}$}}; %E2=

\node[]() at (13,1.775) {{\scriptsize  
$\big\{ a^{1,m_{12}}_{12},a^{2,0}_{12},a^{0,0}_{34}\big\} =\{1,-1,1.0384\}$}}; %E3=

\node[]() at (12.55,2.25) {{\scriptsize 
$\big\{ a^{s_{12},m_{12}}_{12},a^{s_{34},m_{34}}_{34}\big\} =\{1,1\}$}}; %E4=

\node[]() at (13,2.8) {{\scriptsize  
$\big\{ a^{0,0}_{12},a^{1,0}_{34},a^{2,m_{34}}_{34}\big\} =\{1.1923,-1,1\}$}}; %E5=

\node[]() at (13,3.3) {{\scriptsize  
$\big\{ a^{0,0}_{12},a^{1,0}_{34},a^{2,m_{34}}_{34}\big\} =\{1.1153,-1,1\}$}}; %E6=

\node[]() at (13,3.75) {{\scriptsize  
$\big\{ a^{1,0}_{12},a^{2,m_{12}}_{12},a^{0,0}_{34}\big\} =\{-1,1,1.0384\}$}}; %E7=

\node[]() at (12.55,4.2) {{\scriptsize 
$\big\{ a^{s_{12},m_{12}}_{12},a^{s_{34},m_{34}}_{34}\big\} =\{1,1\}$}}; %E8=

\node[]() at (12.55,6) {{\scriptsize 
$\big\{ a^{s_{12},m_{12}}_{12},a^{s_{34},m_{34}}_{34}\big\} =\{1,1\}$}}; %E9=

\node[]() at (12.55,8) {{\scriptsize 
$\big\{ a^{s_{12},m_{12}}_{12},a^{s_{34},m_{34}}_{34}\big\} =\{1,1\}$}}; %E10=
%------------------------------------------------------------------
%-----------------leybal of the spectra----------------------------
\node[]() at (4,9){{\small $\mathrm{Ni_4Mo_{12}}$}}; %Ni_4Mo_12

%--------Insert box with transitions between states----------------
\node[draw,double,rectangle, very thick, fill=white]() at (4.5,6.5) 
{
\begin{tikzpicture}
\node[]() at (1.15,0) {{\small I} 
{\scriptsize $\lvert 1,1,0,0 \rangle \to \lvert 0, 1, 1, \pm1 \rangle$}}; %First

\node[]() at (1.1,0.5) {{\small II} 
{\scriptsize $\lvert 1,1,0,0 \rangle \to \lvert 1, 0, 1, \pm1 \rangle$}}; %Second

\node[]() at (1.05,1) {{\small III} 
{\scriptsize $\lvert 1,1,0,0 \rangle \to \lvert 0, 1,1, 0 \rangle$} \phantom{I}}; %Third

\node[]() at (1.05,1.5) {{\small IV} 
{\scriptsize $\lvert 1,1,1,\pm1 \rangle \to \lvert 0,1,1, 0 \rangle$}}; %red
\end{tikzpicture}
};
%--------------------------------------------------------------------
\end{tikzpicture}
\caption{
Energy level structure and the corresponding transitions of
Ni$_4$Mo$_{12}$. The blue line and
arrows stands for the ground state energy and the ground state excitations, respectively.
The red arrow marks the excited transition and the corresponding initial
level is shown in red. The dashed lines represent the centers of the two bands.
All transitions are denoted with respect to the experimental data reported in Ref. \cite{nehrkorn_inelastic_2010}. }
\label{fig:NickelSpectra}
\end{figure*}
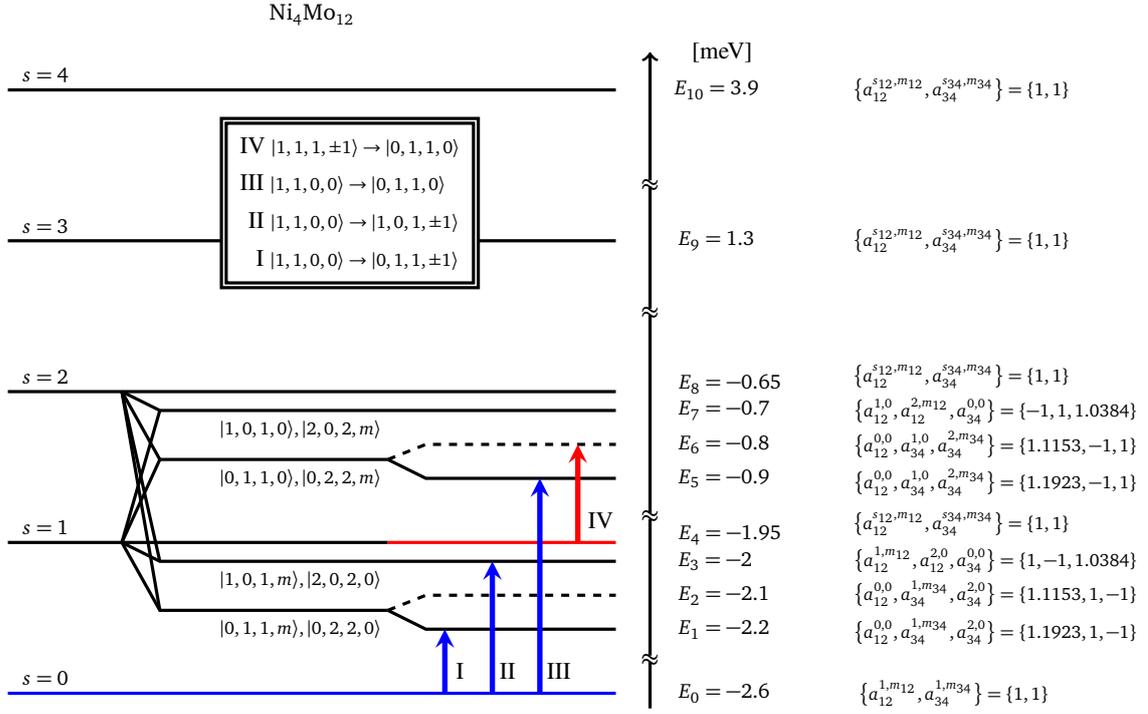

\section{N\lowercase{i}$_4$M\lowercase{o}$_{12}$} \label{sec:nickel}
\subsection{The Hamiltonian}
The indistinguishable spin-one $\mathrm{Ni^{2+}}$ ions of the spin
cluster compound $\mathrm{Ni_4Mo_{12}}$, are arranged
on the vertices of a distorted tetrahedron FIG. \ref{fig:Nickelconf} (a).
The bonds Ni1-Ni2 and Ni3-Ni4 are slightly shorter than the
other four. Distance measurements \cite{furrer_magnetic_PRB_2010}
report a difference of the order of $0.03\mathrm{\mathring{A}}$.

To perform an analysis of the magnetic excitations of the
compound $\mathrm{Ni_4Mo_{12}}$ obtained by INS experiments reported
in Ref. \cite{nehrkorn_inelastic_2010,furrer_magnetic_PRB_2010} we
consider the formalism described in Section \ref{sec:spinmodel}.
According to the symmetry of the magnetic cluster we do the 
imply $J_{ij}=J$ and assume that
the ions Ni1-Ni2 and Ni3-Ni4 are coupled, as shown in FIG.
\ref{fig:Nickelconf} (b) by red lines, which defines
these bonds as intersections of two different planes.
Therefore, we have the total spin eigenstates
$\lvert s_{12},s_{34},s,m \rangle$
four $\sigma$ operators for each constituent magnetic ion and two 
bond operators corresponding to both Ni1-Ni2 and Ni3-Ni4 spin pairs.
The $\sigma$ operators $\hat{\boldsymbol{\sigma}}_1$ and 
$\hat{\boldsymbol{\sigma}}_2$ account for the 
possible changes in the superexchange processes between 
Ni1-Ni2 couple sharing the coefficient 
$a^{s_{12},m_{12}}_{12}$ of the total bond $\sigma$ operator 
$\hat{\boldsymbol{\sigma}}_{12}$. The operators $\hat{\boldsymbol{\sigma}}_3$ 
and $\hat{\boldsymbol{\sigma}}_4$ are associated with the
coefficient $a^{s_{34},m_{34}}_{34}$ of the remaining $\sigma$ operator 
$\hat{\boldsymbol{\sigma}}_{34}$.
Consequently from \eqref{eq:AddHamiltonian} we obtain the Hamiltonian
\begin{align}\label{eq:NickelHamilton}
\hat{\mathcal{H}} & =  J 
\left( 
\hat{\boldsymbol{\sigma}}_1\cdot\hat{\mathbf{s}}_2 + 
\hat{\boldsymbol{\sigma}}_2\cdot\hat{\mathbf{s}}_1 +
\hat{\boldsymbol{\sigma}}_3\cdot\hat{\mathbf{s}}_4 +
\hat{\boldsymbol{\sigma}}_4\cdot\hat{\mathbf{s}}_3
\right) 
\nonumber \\ 
& + J 
\left( 
\hat{\boldsymbol{\sigma}}_{12}\cdot\hat{\mathbf{s}}_{34} +
\hat{\boldsymbol{\sigma}}_{34}\cdot\hat{\mathbf{s}}_{12}
\right).
\end{align}
With the applied effective spin-one spins the tetramer
exhibits in total eighty one eigenstates without counting the
quadrupolar, octupolar and other eigenfunctions related with higher
symmetries.
The ground state of this nanomagnet is a singlet with possible eigenstates
$\{\lvert 0,0,0,0 \rangle,\lvert 1,1,0,0 \rangle, \lvert 2,2,0,0 \rangle\}$.
On the other hand, the selection rules imply that the ground state
excitations must be related with singlet-triplet transitions and
since the quantum numbers $s_{14}$ and $s_{23}$ cannot be
simultaneously varied, we
deduce that the ground state is,
related to the formation of two local triplets, i.e. $s_{14}=1$ and $s_{23}=1$.
The triplet
eigenstates are eighteen. Those, three in total,
%$\{\lvert 2,2,1,\pm1 \rangle, \lvert 2,2,1,0 \rangle\}$
characterized by the local quintets $s_{14}=2$ and $s_{23}=2$
are not adequate to the
established selection rules and nine are identified as connected
to experimental spectra.

\subsection{Energy levels}
According to the selected coupling scheme we denote the
eigenvalues of Hamiltonian in Eq. \eqref{eq:NickelHamilton} by 
$E^{m}_{s_{12},s_{34},s}$.
The ground state is $\lvert 1,1,0,0 \rangle$. Therefore, using 
\eqref{eq:Constraint_ij} we get $a^{1,m_{12}}_{12}=a^{1,m_{34}}_{34}=1$
and taking into account \eqref{eq:NickelHamilton} we obtain
\[
E^{0}_{1,1,0} = -8J.
\]
With the eigenstates when the spins of Ni1 and Ni2 ions
are coupled in a singlet, the parameter $a^{0,0}_{12}$ remains unconstrained 
and can be determined using INS experimental data.  
For the corresponding energy we get
\[
E^{0}_{0,1,1} = - 2Ja^{1,0}_{34}-4Ja^{0,0}_{12},
\quad
E^{\pm 1}_{0,1,1} = - 2Ja^{1,\pm 1}_{34} -4Ja^{0,0}_{12}. 
\]
Analysis of Nickel spectrum yields
$a^{0,0}_{12}=\{c^1_{12},c^2_{12}\}$. Thus
\[
E^{\pm1}_{0,1,1} \in \left\lbrace -2J-4Jc^1_{12}, 
-2J-4Jc^2_{12}\right\rbrace. 
\]
Moreover, when $m_{34}=0$ we have $a^{1,0}_{34}\in\{1,-1\}$, see \eqref{eq:as0_ij}.
Hence
\[
E^{0}_{0,1,1} \in \left\lbrace -2J-4Jc^1_{12}, 
-2J-4Jc^2_{12}, 2J-4Jc^1_{12}, 2J-4Jc^2_{12} \right\rbrace. 
\]
For the eigenstates corresponding to the Ni3-Ni4
singlet bond, the value of $a^{0,0}_{34}$ remains unconstrained 
leading to $m=m_{12}$ and according to \eqref{eq:NickelHamilton}
we have
\[
E^{0}_{1,0,1} = - 2Ja^{1,0}_{12}-4Ja^{0,0}_{34},
\quad
E^{\pm 1}_{1,0,1} = 
- 2Ja^{1,\pm 1}_{12} -4Ja^{0,0}_{34}.  
\]
We found no evidence
that $a^{0,0}_{34}$ should be discrete and we set $a^{0,0}_{34}=c_{34}$.
Further, with $m_{12}=0$ we have $a^{1,0}_{12}\in\{1,-1\}$. 
As a result we get
\[
E^{0}_{1,0,1} \in \left\lbrace - 2J-4Jc_{34}, 2J-4Jc_{34}\right\rbrace,
\quad
E^{\pm 1}_{1,0,1} = -2J-4Jc_{34}. 
\]
For all of the remaining triplets
$\lvert 1,1,1,m\rangle$, $\lvert
2,2,1,m\rangle$, $\lvert
2,1,1,m\rangle$ and $\lvert 1,2,1,m\rangle$
, where $m=0,\pm 1$, the corresponding 
coefficient are constrained
$a^{s_{12},m_{12}}_{12}=1$ and $a^{s_{34},m_{23}}_{34}=1$. 
Thus, we obtain
\[
E^{m}_{1,1,1} = E^{m}_{2,2,1} =
E^{m}_{2,1,1} = E^{m}_{1,2,1} = -6J.
\]
Furthermore, the tetramer exhibits also a singlet bond at the quintet level.
The energies associated with the Ni1-Ni2 bond with singlet eigenstates 
$\lvert 0,2,2,m \rangle$, where $m\equiv m_{34}$ are
\[
\begin{array}{l}
E^{0}_{0,2,2} = 2Ja^{2,0}_{34}-4Ja^{0,0}_{12}, \quad
E^{\pm 1}_{0,2,2} = 2Ja^{2,\pm 1}_{34} -4Ja^{0,0}_{12}, 
\\ [0.2cm]
E^{\pm 2}_{0,2,2} = 2Ja^{2,\pm 2}_{34} -4Ja^{0,0}_{12}.
\end{array}
\]
With $a^{2,0}_{34}\in\{1,-1\}$ and $a^{0,0}_{12}=\{c^1_{12},c^2_{12}\}$
we have
\[
E^{0}_{0,2,2} \in \left\lbrace -2J-4Jc^1_{12}, 
-2J-4Jc^2_{12}, 2J-4Jc^1_{12}, 2J-4Jc^2_{12} \right\rbrace.
\]
When $m_{34}=\pm 1, \pm 2$, we obtain
\[
\begin{array}{c}
E^{\pm 1}_{0,2,2}\in\left\lbrace2J-4Jc^1_{12}, 2J-4Jc^2_{12}\right\rbrace, 
\\ [0.2cm]
E^{\pm 2}_{0,2,2}\in\left\lbrace2J-4Jc^1_{12}, 2J-4Jc^2_{12}\right\rbrace.
\end{array}
\]
Once at the quartet level the spins of third and fourth ions form a singlet, 
where the corresponding eigenstates are 
$\lvert 2,0,2,m \rangle$, then the Hamiltonian in \eqref{eq:NickelHamilton}
yield the following energy values
\[
\begin{array}{l}
E^{0}_{2,0,2} = 2Ja^{2,0}_{12}-4Ja^{0,0}_{34}, \quad
E^{\pm 1}_{2,0,2} = 2Ja^{2,\pm 1}_{12} -4Ja^{0,0}_{34}, 
\\ [0.2cm]
E^{\pm 2}_{2,0,2} = 2Ja^{2,\pm 2}_{12} -4Ja^{0,0}_{34}.
\end{array}
\]
Similarly, taking into account that $a^{2,0}_{12}\in\{1,-1\}$ and 
$a^{0,0}_{34}=c_{34}$, we obtain
\[
\begin{array}{l}
E^{0}_{2,0,2} \in \left\lbrace-2J-4Jc_{34}, 2J-4Jc_{34}\right\rbrace, 
\\ [0.2cm]
E^{\pm 1}_{2,0,2} = 2J-4Jc_{34},
\quad E^{\pm 2}_{2,0,2} = 2J-4Jc_{34}.
\end{array}
\]
For the other twelve quintet states the coefficients
$a^{s_{12},m_{12}}_{12}=a^{s_{34},m_{34}}_{34}=1$. 
Therefore,
\[
E^{m}_{2,2,2} = E^{m}_{1,1,2} =
E^{m}_{2,1,2} = E^{m}_{1,2,2} = -2J.
\]
For the two remaining levels and the corresponding eigenstates, we obtain 
$a^{s_{12},m_{12}}_{12}=a^{s_{34},m_{34}}_{34}=1$.
The energy sequence follows the Land\'e interval rule $E_{s+1} -
E_{s} = 2Js$, see FIG. \ref{fig:NickelSpectra}. 
The septet level is twenty one fold degenerate. It is defined by the
vectors $\lvert 2,1,3,m \rangle, \lvert 1,2,3,m \rangle, \lvert
2,2,3,m \rangle$ with $m=0,\pm1,\pm2,\pm3$. All corresponding
energies are equal
\[
E^{m}_{2,1,3} = E^{m}_{1,2,3} =
E^{m}_{2,2,3} = 4J.
\]
For the nonet state $\lvert 2,2,4,m \rangle$,
where $m=0,\pm1 ,\pm2,\pm3,\pm4$ we end up with
\[
E^{m}_{2,2,4} = 12J.
\]
The described energy level structure is illustrated on 
FIG. \ref{fig:NickelSpectra}.	
In what follows we find the following notations more convenient
\begin{tabular}{lll}
$E_0 = -8J,\ $ & $E_1 = -2J-4Jc^1_{12},\ $ & $E_2 = -2J-4Jc^2_{12},$ \\
$E_3 = -2J-4Jc_{34},\ $ & $E_4=-6J,\ $ & $E_5 = 2J-4Jc^1_{12}$ \\
$E_6 = 2J-4Jc^2_{12},\ $ & $E_7 = 2J-4Jc_{34},\ $ & $E_8=-2J,$ \\
$E_9=4J,\ $ & $E_{10}=12J$. & ${}$
\end{tabular}
		
\subsection{Scattering Intensities}

The INS selection rules are $\Delta s = 0, \pm 1$, $\Delta
m = 0, \pm 1$ and $\Delta s_{12} = 0, \pm 1$, $\Delta s_{34} = 0, \pm
1$. Here the transitions $\Delta s_{12} \ne 0$
and $\Delta s_{34} \ne 0$ are not allowed simultaneously.

\begin{figure}[th!]
\centering
\includegraphics[scale=0.9]{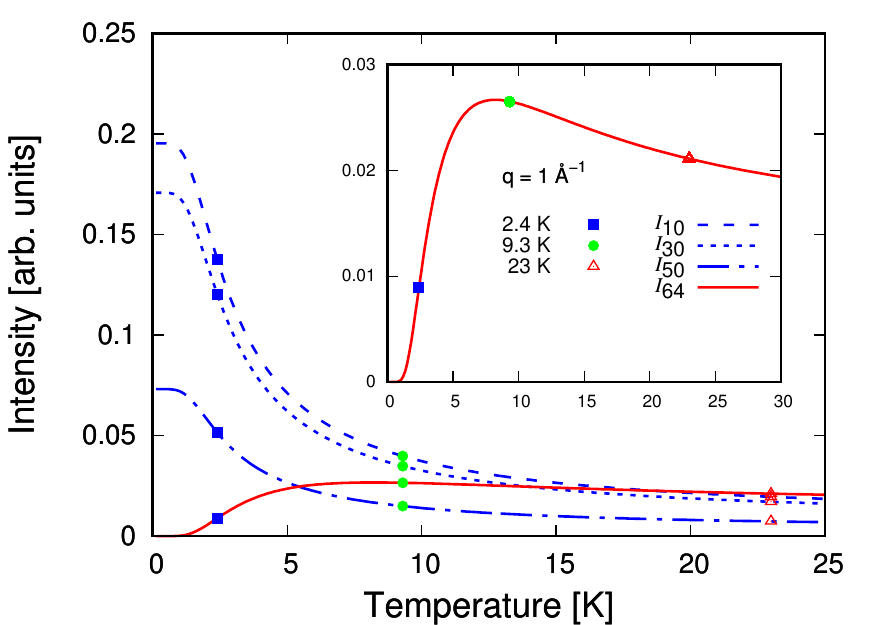}
\caption{
Intensities as a function of the absolute temperature. $I_{10}$, $I_{30}$
and $I_{50}$ correspond to the ground state transitions with
energies 0.4 meV, 0.6 meV and 1.7 meV, respectively. The
intensity $I_{64}$ in the inset stands for the excited transition with
energy 1.15 meV.
The blue squares, the green cirlces and red triangles point to the
values of intensities in TAB. \ref{tab:NickelIntensThreeTemp}.
\label{fig:NiIntensity}
}
\end{figure}

\begin{figure}[th!]
\centering
\includegraphics[scale=0.9]{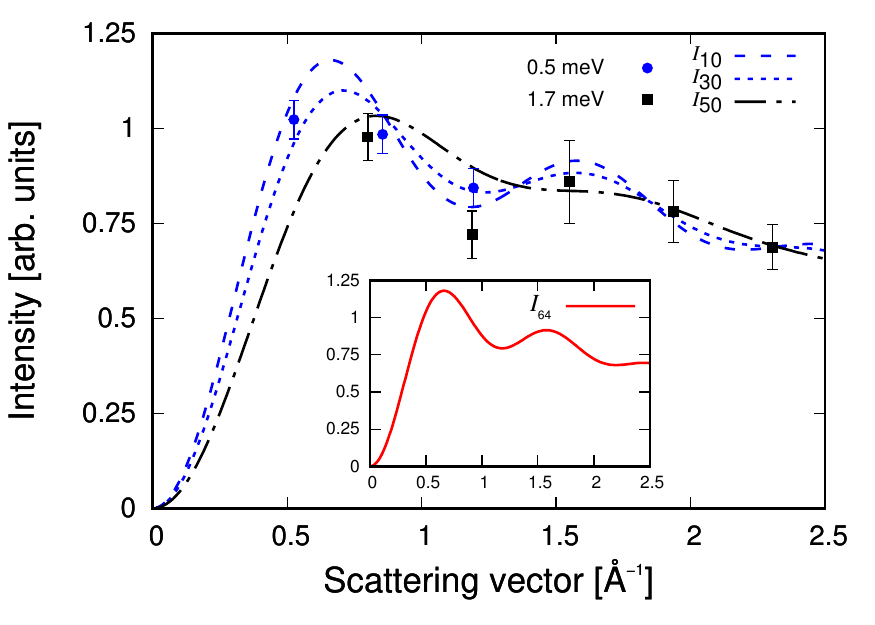}
\caption{
Normalized by $\gamma_{n'n}$ intensities as a function of the
scattering vector along with the experimental data of Ref.
\cite{furrer_magnetic_PRB_2010}. $I_{10}$, $I_{30}$ and $I_{50}$ correspond
to ground state transitions with energies 0.4 meV, 0.6 meV and
1.7 meV, respectively. The intensity $I_{64}$ stands for the
excited transition with energy 1.15 meV. The inset
shows the intensity $I_{64}$ that coincides
with the function $I_{10}$.
\label{fig:NiIntFFactor}
}
\end{figure}

Using \eqref{eq:ScatterFunctions} we obtain $S^{\alpha
\beta}(\mathbf{q},\omega_{n'n})+S^{\beta
\alpha}(\mathbf{q},\omega_{n'n})=0, \ \forall \ n,n'$ and $\alpha \ne
\beta$. 
The analysis of the scattering intensities reveals the experimental magnetic
excitation at $0.4$ meV
\cite{nehrkorn_inelastic_2010,furrer_magnetic_PRB_2010} corresponding
to the transition between the ground state and the singlet state 
$\lvert 0,1,1,\pm1 \rangle$ with
\begin{subequations}\label{eq:ScatteringNikel12States1}
\begin{align}
S^{\alpha \alpha}(\mathbf{q},\omega_{10}) & = \tfrac{4}{9} 
[1-\cos(\mathbf{q}\cdot\mathbf{r}_{12})]p_0,
\\
S^{zz}(\mathbf{q},\omega_{10}) & = 0,
\end{align}
\end{subequations}
where $\alpha = x,y$. 
The magnetic excitation at $0.6$ meV
\cite{nehrkorn_inelastic_2010,furrer_magnetic_PRB_2010} is associated
with the eigenstate $\lvert 1,0,1,\pm1 \rangle$ and the scattering
functions
\begin{subequations}\label{eq:ScatteringNikel34States1}
\begin{align}
S^{\alpha \alpha}(\mathbf{q},\omega_{30}) & = 
\tfrac{4}{9} [1-\cos(\mathbf{q}\cdot\mathbf{r}_{34}) ]p_0,
\\
S^{zz}(\mathbf{q},\omega_{30}) & = 0,
\end{align}
\end{subequations}
where $\alpha = x,y$. The functions
\eqref{eq:ScatteringNikel12States1} differ from
\eqref{eq:ScatteringNikel34States1} due to the spatial orientations
of the spin bonds with
$\mathbf{r}_{12}\cdot\mathbf{r}_{34} = 0$. 
For the same reason, we
deduce that the third cold peak at $1.7$ meV \cite{nehrkorn_inelastic_2010,furrer_magnetic_PRB_2010} is 
related with the transition between the ground state
and non magnetic triplet $\lvert 0,1,1,0 \rangle$.
For $\alpha = x,y$ the corresponding
scattering functions are
\begin{align*}
S^{zz}(\mathbf{q},\omega_{50}) & = 
\tfrac{4}{9} [1-\cos(\mathbf{q}\cdot\mathbf{r}_{12}) ]p_0,
\\
S^{\alpha \alpha}(\mathbf{q},\omega_{50}) & = 0.
\end{align*}
The excited magnetic transition at around $1.2$ meV \cite{nehrkorn_inelastic_2010,furrer_magnetic_PRB_2010} is 
nicely reproduced by the scattering functions
\begin{align*}
S^{\alpha \alpha}(\mathbf{q},\omega_{64}) & = 
\tfrac{2}{3} [1-\cos(\mathbf{q}\cdot\mathbf{r}_{12})] p_4,
\\
S^{zz}(\mathbf{q},\omega_{64}) & = 0,
\end{align*}
where $\alpha = x,y$. The initial state is
%\eqref{eq:Nickel2BondTripletState}
given by the triplet state $\lvert 1,1,1,\pm1 \rangle$ with two triplet bonds
and the final one appears to be $\lvert 0,1,1,0 \rangle$.
Hence if the neutron scatters from the Ni3-Ni4 dimer, then we have
$\mathbf{q}\cdot\mathbf{r}_{12} = 0$ and
$\mathbf{q}\cdot\mathbf{r}_{34} > 0$. 
We remark that the orthogonality
of $\mathbf{r}_{12}$ and $\mathbf{r}_{34}$ can be considered
independently from the formalism presented in Section \ref{sec:spinmodel}.
Nevertheless, with the coefficients $a^{s_{12},m_{12}}_{12}$ and 
$a^{s_{34},m_{34}}_{34}$ one can uniquely identify the 
two spin bonds and distinguish $I_{10}$ from $I_{30}$. 
Moreover, one can distinguish the eigenvalues
{of tetramer Hamiltonian} corresponding to $m=0$ and $m\ne0$, 
with $S^{zz}(\mathbf{q},\omega_{n'n}) = 0$ and 
$S^{xx}(\mathbf{q},\omega_{n'n})=0$, $S^{yy}(\mathbf{q},\omega_{n'n})=0$, 
respectively.
This affects directly the integrated intensities, such that choosing
$\mathbf{r}_{12} = (0,0,r^z)$ and $\mathbf{r}_{34} = (r^x,0,0)$ from
\eqref{eq:ScateringIntensities} yields
\begin{equation*}\label{eq:NickelIntegIntensities}
\begin{array}{l}
\displaystyle
I_{10} \propto \gamma_{10}
	\left[ 
	1-\frac{\sin(\mathrm{q}\mathrm{r})}{\mathrm{q}\mathrm{r}}
	\right]  F^2(\mathrm{q}),
\\ [0.5cm]
\displaystyle
	I_{30} \propto \gamma_{30} %\frac{20}{27}p_0 
	\left[ 
		1 - 6\frac{\sin(\mathrm{q}\mathrm{r})}{5(\mathrm{q}\mathrm{r})^3} - 3\frac{\sin(\mathrm{q}\mathrm{r})}{5\mathrm{q}\mathrm{r}} + 6\frac{\cos(\mathrm{q}\mathrm{r})}{5(\mathrm{q}\mathrm{r})^2}
	\right] F^2(\mathrm{q}),
\\[0.5cm]
\displaystyle
	I_{50} \propto \gamma_{50} %\frac{8}{27}p_0
	\left[ 
	1-3\frac{\sin(\mathrm{q}\mathrm{r})}{(\mathrm{q}\mathrm{r})^3}+
	3\frac{\cos(\mathrm{q}\mathrm{r})}{(\mathrm{q}\mathrm{r})^2}
	\right] F^2(\mathrm{q}),
\\[0.5cm]
\displaystyle
	I_{64} \propto \gamma_{64}  %\frac{4}{3}p_3
	\left[ 
	1-\frac{\sin(\mathrm{q}\mathrm{r})}{\mathrm{q}\mathrm{r}}
	\right]  F^2(\mathrm{q}),
\end{array}
\end{equation*}
where 
\[
\gamma_{10} = \tfrac{8}{9}p_0, \ \gamma_{30} = \tfrac{20}{27}p_0 \ 
\gamma_{50} = \tfrac{8}{27}p_0, \ \gamma_{64} = \tfrac{4}{3}p_4,
\]
and $r=|\mathbf{r}_{12}|=|\mathbf{r}_{34}|$. 
The integrated intensities as a function of temperature
are shown on FIG. \ref{fig:NiIntensity}. According to Ref.
\cite{furrer_magnetic_PRB_2010} the average distance between
Ni-Ni ions is $\mathrm{r}=6.68 \ \mathrm{\mathring{A}}$. 
The magnitude of the scattering vector is fixed at $\mathrm{q}=1 \
\mathrm{\mathring{A}}^{-1}$ and the form factor is given by 
\eqref{eq:MagFormFactorIntegrate}. 
The dependence of normalized intensities, $I_{n'n} \to
I_{n'n}/\gamma_{n'n}$, on the scattering vector is shown on FIG.
\ref{fig:NiIntFFactor}.
%with some experimental values given in TAB.
%\ref{tab:NickelIntensThreeTemp}.
	
\begin{table}[!ht]
\caption{
Calculated values of integrated intensities $I_{n'n}$ [arb. units] at
temperatures 2.4, 9.3 and 23 K, shown on FIG.
\ref{fig:NiIntensity} as blue squares, green circles and red trianlges,
respectively. \label{tab:NickelIntensThreeTemp}
}
\begin{ruledtabular}%	\toprule[1.2pt]
\begin{tabular}{lcccc}
Transitions & I & II & III & IV  \\
$T$ [K] & $I_{10}$ & $I_{30}$ & $I_{50}$ & $I_{64}$  \\
\hline %		\midrule[0.5pt]	
2.4 & 0.137(6) & 0.120(3) & 0.051(5) & 0.008(9)  \\
9.3 & 0.039(8) & 0.034(8) & 0.014(9) & 0.026(5)  \\
23 & 0.019(5) & 0.017(1) & 0.007(3) & 0.021(1)  \\	
		\end{tabular}
	\end{ruledtabular}%	\toprule[1.2pt]
\end{table}	

\begin{table}[!h]
	\caption{
		Values of the coupling constants and the quantities $a^{0,0}_{12}$, 
		$a^{0,0}_{34}$
		for all magnetic excitations with energies $E_{n'n}$ obtained
		by taking into account the experimental data of Ref.
		\cite{nehrkorn_inelastic_2010,furrer_magnetic_PRB_2010}.
		\label{tab:Nickel}
	}
	\begin{ruledtabular}%	\toprule[1.2pt]
		\begin{tabular}{lcccc}
			Transitions & I & II & III & IV  \\
			$\mathrm{Ni_4Mo_{12}}$ & $E_{10}$ & $E_{30}$ & $E_{50}$ & $E_{64}$  \\
			\hline %		\midrule[0.5pt]	
			$E_{n'n}$[meV] & 0.4 & 0.6 & 1.7 & 1.15  \\
			$J_{\phantom{c^1_{12}}}$ [meV] & 0.325 & 0.325 & 0.325 & 0.325  \\
			$J_{c^1_{12}}$ [meV] & 0.372 & $-$ & 0.372 & $-$  \\
			$J_{c^2_{12}}$ [meV] & $-$ & $-$ & $-$ & 0.353  \\
			$J_{c_{34}}$ [meV] & $-$ & 0.334 & $-$ & $-$  \\
			$c^1_{12}$ & 1.1923 & $-$ & 1.1923 & $-$ \\
			$c^2_{12}$ & $-$ & $-$ & $-$ & 1.1153 \\
			$c_{34}$ & $-$ & 1.0384 & $-$ & $-$ \\ 
		\end{tabular}
	\end{ruledtabular}%	\toprule[1.2pt]
\end{table}	
		
\subsection{Energy of the magnetic transitions}
The energy transition $E_{ij}$ between $i$th and $j$th levels, 
corresponding to the calculated scattering intensities are
\begin{align*}
	E_{10} & = 6J-4Jc^1_{12},
\qquad
	E_{30} = 6J-4Jc_{34},
\\
	E_{50} & = 10J-4Jc^1_{12},
\qquad
	E_{64} = 8J-4Jc^2_{12}.
\end{align*}
From the last equations we can take advantage of one more constraint to
determine $J$, $E_{50}-E_{10} = 4J$. According to the
experimental data
\cite{nehrkorn_inelastic_2010,furrer_magnetic_PRB_2010} the ground
state magnetic excitations are grouped in two relatively broadened
peaks. The first peak is centred at about 0.5 meV and the second one at 1.7
meV. 
Furthermore, the first peak is composed of two subbands with energies
$E_{10}=0.4$ meV and $E_{30}=0.6$ meV. The width of the second peak can be
explained by the presence of an energy band, where the transition
energies are restricted in the region 1.6 meV to 1.8 meV. 
Therefore,
setting $E_{50} = 1.7$ meV we obtain $E_{50}-E_{10} = 1.3$ meV and $J
= 0.325$ meV. The computed energy
transitions are depicted on FIG. \ref{fig:NickelSpectra}. The
centers of both energy bands referring
to the value $c^{2}_{12} = 1.1153$ are shown by 
dashed lines. The energies of all transitions and
the corresponding parameters are given in TAB. \ref{tab:Nickel}.

\section{Conclusion}\label{sec:conclusion}
We propose a formalism that introduces
a systematic
approach for exploring the physical properties
of molecular magnets.
The underlying concept lies on the hypothesis
that due the cluster symmetry, as well as its shape,
size 
and the chemical structure that surrounds 
the magnetic ions, the exchange pathway between two particular metal ions 
is not unique leading to a variation of the relevant exchange energy.
  
To check the validity of this hypothesis
we construct Hamiltonian \eqref{eq:AddHamiltonian} 
that accounts for discrete coupling parameters
derived via the relations
\eqref{eq:Sigma_i}, \eqref{eq:Sigma_ij} and \eqref{eq:Sigma_k}
that allows one to distinguish spin coupling mechanisms among equivalent
magnetic ions. 

We apply this formalism to explore the magnetic excitations of 
the compounds $\mathrm{A_3Cu_3(PO_4)_4}$ with (A = Ca, Sr, Pb) 
and $\mathrm{Ni_4Mo_{12}}$ obtaining results consistent with INS experiments
\cite{matsuda_magnetic_2005,podlesnyak_magnetic_2007} and
\cite{nehrkorn_inelastic_2010,furrer_magnetic_PRB_2010}, respectively.
%
%--------------------A3Cu3(PO4)4-------------------------------
We deduce that the ground state energy of the 
trimers $\mathrm{A_3Cu_3(PO_4)_4}$ (A = Ca, Sr, Pb) is associated 
with the Cu1-Cu3 triplet bond. We obtained a thin energy band composed 
of two very close energy levels corresponding to the Cu1-Cu3 singlet
(see e.g. FIG \ref{fig:CopperEnergyStructure}).
The neutron energy loss associated with the first and the excited 
spin excitations is due to the transitions from triplet to singlet 
Cu1-Cu3 state. The second ground state excitation is the result of 
doublet-quartet transitions.
Further, the discrete parameter $a^{0,0}_{13}\in\{c^1_{13},c^2_{13}\}$, with
$|c^i_{13}|<1$ for $i=1,2$, shows that in the doublet level characterized by 
eigenstate $\lvert 0,\tfrac12, \pm\tfrac12\rangle$ the field along all bridges 
between edge ions have less strength and the exchange could not be 
maintained. 
Thus, according to our calculations the next-nearest neighbor coupling 
	$J_{13}\in\{J_{c^1_{13}},J_{c^2_{13}}\}$ is negligible, see TAB. \ref{tab:Copper}.
The value $|c^1_{13}-c^2_{13}| = 0.031$ signals for the small 
variations of the {next-nearest neighbor exchange coupling} 
which therefore explains the
sharpness of the experimentally observed peaks 
\cite{matsuda_magnetic_2005,podlesnyak_magnetic_2007}.

%----------------------Ni_4Mo12---------------------------
Studying the INS spectra of the compound $\mathrm{Ni_4Mo_{12}}$ 
with the proposed in Sec. \ref{sec:spinmodel} approach we were
able to derive a detailed picture for the neutron scattering 
intensities FIGs. \ref{fig:NiIntensity} and \ref{fig:NiIntFFactor}.
Hamiltonian \eqref{eq:NickelHamilton} leads to
 energy spectrum with two energy bands, 
shown in FIG. \ref{fig:NickelSpectra}. These bands are related to 
the fact that the tetramer cluster exhibits two distinguishable 
with respect to the coefficients $a^{s_{12},m_{12}}_{12}$ and 
$a^{s_{34},m_{34}}_{34}$ bonds. 
We ascribe this feature to the
difference in the chemical environment around Ni1-Ni2 and Ni3-Ni4 couples.
This allowed a unique identification of the magnetic 
excitations. Thereby, the obtained energy bands explain the width of second 
ground state peaks centred at 1.7 meV and the splitting of the first 
one centred at 0.5 meV. The splitting was found to be the consequence 
of the different spatial orientation of the Ni1-Ni2 and Ni3-Ni4 bonds 
(see e.g. FIG. \ref{fig:Nickelconf}). 
In particular, for $s_{12} = 0$, $s_{34} = 0$ and $i=1,2$ 
we get $|c^i_{12}|>1$ and 
$|c_{34}|>1$, respectively. 
Besides, according to \eqref{eq:cij_a1} we have $J<J_{c^{i}_{12}}$ and $J<J_{c_{34}}$,
	see TAB. \ref{tab:Nickel}.
These inequalities signals that the strength 
of the exchange is amplified.
Furthermore, the inequality $J_{c_{34}}<J_{c^{i}_{12}}$
indicates that most probably
the field has less strength along Ni3-Ni4 bond than the Ni1-Ni2 one.

%------------------furder results--------------------------
In the present study we confined ourselves to the explanation of
experimental INS spectra of some representative trimers and tetramers.
We would like to mention that the method can be applied to other
magnetic properties, such as the magnetization and the susceptibility.
We would like to anticipate that preliminary results are encouraging
and will be the subject of a separate paper.

\begin{acknowledgments}
The authors are indebted to Prof. N.S. Tonchev, Prof. N. Ivanov and
Prof. J. Schnack for very helpful discussions, and 
to Prof. M. Matsuda for providing us with the experimental data used
in FIGs. \ref{PbIntensity} and \ref{fig:CuFFIntensity}.
This work was supported by the Bulgarian National Science Fund under
contract DN/08/18.
\end{acknowledgments}

%\bibliography{molmag}

%
\end{document}